\documentclass{aastex}
\usepackage{emulateapj5}
\usepackage{apjfonts}
\usepackage{epsf}
\usepackage{graphicx}

\makeatletter

\newenvironment{inlinefigure}{%
\def\@captype{figure}%
\noindent\begin{minipage}{0.999\linewidth}\begin{center}}
{\end{center}\end{minipage}\smallskip}
\makeatother

\lefthead{NAGASHIMA}
\righthead{MISSING LINK IN THE PRESS-SCHECHTER FORMALISM}

\newcommand{\eikx}{{e^{-i{\bf k\cdot x}}}}
\newcommand{\tW}{{\tilde{W}}}
\newcommand{\bfeta}{\mbox{\mathversion{bold}$\eta$\mathversion{normal}}}
\newcommand{\bfzeta}{\mbox{\mathversion{bold}$\zeta$\mathversion{normal}}}
\begin{document}

\submitted{\rm\it NAOJ-Th-Ap2001, No.44}

\title{A Solution to the Missing Link in the Press-Schechter Formalism}
\author{Masahiro Nagashima}
\affil{Division of Theoretical Astrophysics, 
National Astronomical Observatory,
Mitaka, Tokyo 181-8588, Japan;
\email{masa@th.nao.ac.jp}}

\begin{abstract}
 The Press-Schechter (PS) formalism for mass functions of the
 gravitationally collapsed objects is reanalyzed.  It has been suggested
 by many authors that the PS mass function agrees well with the mass
 function given by $N$-body simulations, while many too simple
 assumptions are contained in the formalism.  In order to understand why
 the PS formalism works well we consider the following three effects on
 the mass function: the filtering effect of the density fluctuation
 field, the peak ansatz in which objects collapse around density maxima,
 and the spatial correlation of density contrasts within a region of a
 collapsing halo because objects have non-zero finite volume.  According
 to the method given by Yano, Nagashima \& Gouda (YNG) who used an
 integral formula proposed by Jedamzik, we resolve the missing link in
 the PS formalism, taking into account the three effects.  While YNG
 showed that the effect of the spatial correlation alters the original
 PS mass function, in this paper, we show that the filtering effect
 almost cancels out the effect of the spatial correlation and that the
 original PS is almost recovered by combining these two effects,
 particularly in the case of the top-hat filter.  On the other hand, as
 for the peak ansatz, the resultant mass functions are changed
 dramatically and not in agreement especially at low-mass tails in the
 cases of the Gaussian and the sharp $k$-space filters.  We show that
 these properties of the mass function can be interpreted in terms of
 the kernel probability $P(M_{1}\vert M_{2})$ in the integral formula
 qualitatively.
\end{abstract}


\section{Introduction}
It is one of the most important work to determine the number density of
collapsed objects from the statistical properties of the density
fluctuation field at an early stage of cosmic structure evolution.  Such
number density, or the {\it mass function of halos}, constitutes a main
body of analyses of cosmic structures such as galaxies.  The pioneering
work was done by Press \& Schechter (1974, hereafter PS).  They derived
a mass function by relating a one-point distribution function of density
fluctuation with a multiplicity function of halos.  It is very simple
and useful, and therefore has been applied to many problems related with
statistics of cosmic objects by many authors.  Moreover, in order to
take into account the formation histories of galaxies, it has been
extended to estimate the mass function of progenitors of a present-day
halo (Bower 1991; Bond et al. 1991, hereafter BCEK; Lacey \& Cole 1993,
hereafter LC; Kauffmann \& White 1993; Somerville \& Kolatt 1998).  Thus
to remove uncertainties in the PS formalism and to refine it are very
important.

Recent understanding on the cosmological structure formation is based on
a cold dark matter (CDM) model which is a kind of the hierarchical
clustering scenario.  In the CDM model, the Universe is dominated by a
cold dark matter gravitationally, and the self-gravitating structures
are formed via a gravitational growth of small initial density
fluctuations.  Since higher density regions collapses earlier and since
the power at smaller scales of the fluctuations is stronger than that at
larger scales, larger objects which are formed by collapse of
large-scale fluctuations are formed by clustering of smaller objects
hierarchically.  Formation processes of objects such as galaxies and
galaxy clusters are qualitatively well understood in the context of such
hierarchical clustering scenario.  However, once we would like
quantitative estimate, especially in regard with the statistical
properties such as luminosity function of galaxies and abundance of
clusters of galaxies, we face on difficulties caused by uncertainties to
estimate the mass function of cosmological objects.

The PS formalism is the most useful even at present.  This formula
assumes a Gaussian distribution of density contrast and spherically
symmetric collapse of objects (briefly reviewed in \S 2).  In the PS
formalism, however, there is a problem of the so-called
``fudge-factor-of-two''.  Because their consideration was only in
overdense regions, they simply multiplied the factor of two in order to
consider all matter in the Universe.

Another approach explaining the distribution of cosmological structures
has been developed, that is, the peak formalism.  This was firstly
investigated by Doroshkevich (1970), and extensively studied by Peacock
\& Heavens (1985) and Bardeen et al. (1986, hereafter BBKS).  In this
formula, considering configurations of local density maxima through the
first and second derivatives of the density fluctuation, the number
density of the density peaks is derived.  However, peaks in a field with
larger smoothing scale often include those with smaller smoothing scale.
Thus the mass function derived from the peak formalism is in agreement
with numerical experiments only at very large mass scales (Appel \&
Jones 1990).  In order to derive a correct mass function, we must remove
such small scale peaks contained in larger peaks (the so-called {\it
cloud-in-cloud problem}).  By using ``the confluent system formalism'',
Manrique \& Salvador-Sol{\'e} (1995) derived a mass function taking into
account the removal of nested peaks in the case of a Gaussian filter.

Peacock \& Heavens (1990; hereafter PH90) and BCEK attempted to solve
the cloud-in-cloud problem by considering a trajectory of a density
fluctuation against the varying smoothing scale.  The method is called
the {\it excursion set formalism} according to BCEK.  Assuming the
spherical collapse, the threshold for the collapse is determined by only
the value of the density fluctuation, independent of mass scale.  Thus
we adopt as a mass scale of the density fluctuation of a viriallized
object, a maximum mass scale at which the density contrast is firstly
{\it upcrossing} the threshold for decreasing mass scale.  When the
sharp $k$-space filter is adopted, they obtained the fudge-factor-of-two
because of the Brownian motion of the trajectories due to
uncorrelativity of phases of Fourier modes.  However in the case of
general filters, since the correlation between different Fourier modes
originates the non-Markov behavior, a Monte Carlo method or complicated
approximations are required, as done by PH90 and BCEK.  It should be
noted that this problem had already been considered by Epstein (1983;
1984) in the case of a Poisson fluctuation of point masses.

In contrast to the excursion set formalism, which is based on a
differential formula such as the diffusion equation, Jedamzik (1995)
proposed another approach based on an integral equation.  YNG fixed an
inconsistency of estimation of probability in the original analysis by
Jedamzik (1995) and also obtained the fudge-factor-of-two in the case of
the sharp $k$-space filter.  While within the case of the sharp
$k$-space filter, they extended the Jedamzik's integral formula to
include the effect of the spatial correlation which should be considered
within a collapsing region in order to solve the cloud-in-cloud problem.
They found that the factor of two is lost by the spatial correlation and
predicted larger number density of halos compared to the PS mass
function at large scale at which the variance of the density fluctuation
becomes below unity.  They also connected the peak formalism with the
Jedamzik formula and found that the low-mass tail of mass function is
changed.  Nagashima \& Gouda (1997) confirmed that the mass function
given by YNG is in agreement with that given by the Merging Cell model
proposed by Rodrigues \& Thomas (1996), in which the density fluctuation
field is realized by Monte Carlo method and halos are identified
according to the linear theory and the spherical collapse model.

On the other hand, Monaco (1995; 1997a, b;1998) investigated the
dynamical effect of the collapse by using nonlinear dynamics such as the
Zel'dovich approximation (Zel'dovich 1970).  As well known, the critical
density contrast is 1.69 independent of mass scale for the spherical
collapse in the Einstein-de Sitter universe.  However, the probability
finding objects collapsing spherically is very small and most of halos
collapse non-spherically.  Taking into account such direction-dependent
dynamics, if we interpret a first shell-crossing epoch as a collapsing
epoch, it was found that the resultant mass function is similar to the
PS mass function with effectively decreased critical density below 1.69.

The purpose of this paper is to clarify why the PS mass function works
well and well reproduces mass functions obtained by $N$-body simulations
(e.g., Sheth \& Tormen 1999; Jenkins et al. 2001).  Hereafter we call
this problem the {\it missing link in the PS formalism}.  In this paper
we consider the following three effects based on the integral formula
developed by Jedamzik (1995) and YNG.  First the filtering effect is
investigated.  We use the sharp $k$-space, the Gaussian, and the top-hat
filters.  Second the peak effect is considered, based on the {\it peak
ansatz} in which objects collapse around density maxima.  Third the
spatial correlation is introduced in the same way as YNG, but for the
above three filters.  Finally all the effects are considered
simultaneously.  Since it is very complicated that the dynamical effect
as considered by Monaco is introduced simultaneously, we will only
discuss this effect in \S 6.

The structure of this paper is as follows.  First in \S 2, we define the
smoothed density field introducing the {\it filters}.  In \S
\ref{sec:PS}, we review the PS formalism.  In \S \ref{sec:form}, we show
the formulation by using an integral equation, based on Jedamzik (1995)
and YNG.  In \S \ref{sec:result}, we show the resulting mass functions
and consider the effects of the filters, peaks, and spatial correlation.
In \S 6 we discuss some uncertainties which we do not investigate in
detail.  We devote \S \ref{sec:conc} to summary and conclusions.  Many
detailed definitions and equations are summarized in Appendix.

\section{Smoothed Gaussian Random Field}\label{sec:grf}
First, we define the density contrast as $\delta ({\bf x}) \equiv
\frac{\rho ({\bf x})-\bar\rho}{\bar\rho}$, where $\rho({\bf x})$ is
the density at the point ${\bf x}$ and $\bar\rho$ is the cosmic mean
density.  The Fourier mode of the density contrast, $\delta_{\bf k}$,
is obtained by
\begin{equation}
\delta_{\bf k}=\int\delta({\bf x})e^{i{\bf k\cdot x}}d{\bf x}.
\end{equation}
This density contrast is smoothed out by a window function $W_{M}(r)$
with a smoothing mass-scale $M$,
\begin{eqnarray}
\delta_{M}({\bf x})&\equiv&\int W_{M}(\vert{\bf x'-x}\vert)
 \delta({\bf x'})d{\bf  x'}\\
&=&\frac{1}{(2\pi)^{3}}\int\tW(kR)\delta_{\bf k}\eikx d{\bf k}\label{eq:del},
\end{eqnarray}
where $\tW(kR)$ is the Fourier transform of the window function,
$W_{M}(r)$, $R$ is a scale related to the mass $M$, and
$k\equiv\vert{\bf k}\vert$.

The following three filters are often used as the window function by
using the smoothing length $R$:
\begin{enumerate}
\item Top-hat filter\\
\begin{eqnarray}
W_{M}(r)&=&\frac{3}{4\pi R^{3}}\theta(1-\frac{r}{R}),\\
\tW(kR)&=&\frac{3}{(kR)^{3}}(\sin kR-kR\cos kR).
\end{eqnarray}
\item Gaussian filter\\
\begin{eqnarray}
W_{M}(r)&=&\frac{1}{(2\pi
  R^{2})^{3/2}}\exp\left(-\frac{r^{2}}{2R^{2}}\right),\\
\tW(kR)&=&\exp\left(-\frac{k^{2}R^{2}}{2}\right).
\end{eqnarray}
\item Sharp $k$-space filter\\
\begin{eqnarray}
W_{M}(r)&=&\frac{\sin k_{c}r -k_{c}r\cos k_{c}r}{2\pi^{2}r^{3}},\\
\tW(kR)&=&\theta(k_{c}-k),
\end{eqnarray}
\end{enumerate}
where $k_{c}$ is the cut-off wave number, $k_{c}\simeq R^{-1}$, and
$\theta(x)$ is the Heaviside step function.

Next we define a variance of the density field with a smoothing
mass-scale $M$, $\sigma^{2}_{0}(M)$.  The variance is obtained by using
the power spectrum $P(k)\equiv\langle\vert\delta_{\bf
k}\vert^{2}\rangle$ as follows,
\begin{equation}
\sigma^{2}_{0}(M)\equiv\langle\delta_{M}^{2}\rangle=
\frac{1}{(2\pi)^{3}}\int\tW^{2}(kR)P(k)4\pi k^{2}dk.
\end{equation}
The one-point distribution function in the Gaussian random field is
characterized by the above variance.  The normalization of the power
spectrum is given by a characteristic mass $M_{*}$ at which
$\sigma_{0}(M_{*})=1$.

In connection with the definition of $k_{c}$, here we define the mass
$M$ with a scalelength $R$ or $k_{c}$.  In the case of the top-hat
filter, the mass $M$ is clearly related to $4\pi R^{3}\bar{\rho}/3$.  On
the other hand, for the other filters, the relation is not trivial.  For
example, LC proposed that the mass should be defined as the enclosed
mass by the window function $W_{M}(r)$ renormalized to $W_{M}(r=0)=1$.
It should be noted that this uncertainty does not emerge in the case of
the Gaussian filter because it can be renormalized into $M_{*}$ in the
case of the scale-free power spectrum.  As for the sharp $k$-space
filter, in this paper, we adopt a simple relation, $k_{c}R=\pi$, while
LC adopted $k_{c}R=(9\pi/2)^{1/3}$.  The $k_{c}-R$ relation will be
discussed in \S 6.

By using the spherically symmetric collapse model (Tomita 1969; Gunn \&
Gott 1972), the critical density contrast $\delta_{c}$ is equal to 1.69
in the Einstein-de Sitter universe.  So the region with
$\delta\geq\delta_{c}$ is already included in a collapsed region.  We
adopt this value throughout this paper.  Note that Monaco (1995) found
that effective critical density contrast taking into account a
non-spherical collapse is approximated as $\delta_{c}\simeq 1.5$ by
dealing with $\delta_{c}$ as a free parameter and fitting the PS mass
function to that given by him.

In the followings, we restrict ourselves to the Einstein-de Sitter
cosmological model and scale-free power spectra, $P(k)\propto k^{n}$
with $n=0$ and $n=-2$, except for \S\S\ref{sec:CDM} in which we will
show mass functions in a CDM universe.

\section{The Press-Schechter Formalism}\label{sec:PS}
In this section, we review the Press-Schechter formalism briefly.  The
probability of finding the region whose linear density contrast smoothed
on the mass scale $M$, $\delta_{M}$, is greater than or equal to
$\delta_{c}$ is assumed to be expressed by the Gaussian distribution
given by
\begin{equation}   
        f(\geq\delta_{c};M)=\frac{1}{\sqrt{2\pi}\sigma_{0}(M)}   
        \int_{\delta_{c}}^{\infty}\exp\left[-\frac{\delta^{2}}{2\sigma^{2}   
        _{0}(M)}\right]d\delta.   
\label{eqn:gaussdist}
\end{equation}   
This probability corresponds to the volume fraction of the region with
more than or equal to $\delta_{c}$ in the density field with smoothing
scale $M$ to the total volume in a fair sample of the Universe.
Therefore, the difference between $f(\geq\delta_{c};M)$ and
$f(\geq\delta_{c};M+dM)$ represents the volume fraction of the region
for which $\delta_{M}=\delta_{c}$ precisely.  The density contrast of an
{\it isolated} collapsed object must be precisely equal to $\delta_c$
because an object with $\delta>\delta_c$ would be eventually counted as
an object of larger mass scale.  The volume of each object with mass
scale $M$ is $M/{\bar\rho}$. Then we obtain the following relation,
\begin{equation}   
  \frac{Mn(M)}{\bar\rho}dM=-\frac{\partial   
f(\geq\delta_{c};M)}{\partial M}dM,   
\end{equation}      
where $n(M)$ means the number density of objects with mass $M$, that is,
the mass function.  However, the underdense regions are not considered
in the above equation.  Hence, Press and Schechter {\it simply} multiply
the number density by a factor of two,
\begin{equation}   
  \frac{Mn(M)}{\bar\rho}dM=-{\it 2}\frac{\partial f(>\delta_{c};M)}   
  {\partial M}dM.  
\label{eqn:PSform} 
\end{equation}      
This factor of two has long been noted as a weak point in the PS formula
(the so-called cloud-in-cloud problem).  PH90 and BCEK proposed a
solution to this problem by taking account the probability $P_{up}$ that
a subsequent filtering of larger scales results in having $\delta >
\delta_{c}$ at some point, even when at smaller filters, $\delta <
\delta_{c}$ at the same point (the excursion set formalism).  In the
sharp $k$-space filter, because of the uncorrelation between different
Fourier modes $\delta_{\bf k}$, the trajectory of $\delta_{M}$ behaves
as a random walk for adding a new Fourier mode.  In this case, they
found that the factor of two in the PS formula is reproduced.

\section{Integral equation approach}\label{sec:form}
Jedamzik (1995) proposed another approach to the cloud-in-cloud problem
based on an integral equation.
   
Now, we consider the regions whose smoothed linear density contrasts
with smoothing mass scale $M_1$ are above $\delta_{c}$.  Each region
must be included in an isolated collapsed object with mass $M_{2}$
larger than or equal to $M_{1}$.  Therefore, we obtain the following
equation,
\begin{equation}   
  f(\geq\delta_{c};M_{1})=\int_{M_{1}}^{\infty}P(M_{1}\vert M_{2})
   \frac{M_{2}}   
  {\bar\rho}n(M_{2})dM_{2},   
\label{eqn:jed0}
\end{equation}   
where $P(M_{1}\vert M_{2})$ means the conditional probability of finding
a region of mass scale $M_1$ in which $\delta_{M_{1}}$ is greater than
or equal to $\delta_c$, provided it is included in an isolated overdense
region of mass scale $M_2$.\footnote{ In YNG, we used a notation
$P(M_{1},M_{2})$ according to Jedamzik (1995).  In this paper, we adopt
a form $P(M_{1}\vert M_{2})$ in order to stress that this shows a
conditional probability for $\delta_{1}$ above $\delta_{c}$ in a
Gaussian field $M_{1}$ within a collapsing region with $M_{2}$.  }
Hereafter we call the procedure in which the mass functions are
estimated by solving eq.(\ref{eqn:jed0}) the Jedamzik formalism.  If
$P(M_{1}\vert M_{2})$ is simply given by the conditional probability
$p(\delta_{M_{1}}\geq\delta_{c}\vert \delta_{M_{2}}=\delta_{c})$,
$P(M_1\vert M_2)$ is written as follows by using the Bayes' theorem,
\begin{eqnarray}
P(M_{1}\vert M_{2})&=&p(\delta_{M_{1}}\geq\delta_{c}\vert \delta_{M_{2}}=\delta_{c})\nonumber\\
&=&\frac{1}{\sqrt{2\pi}\sigma_{{\rm sub}}}\int
        _{\delta_{c}}^{\infty}\exp\left\{-\frac{1}{2}\frac{(\delta_{M_{1}}-
        \delta_{c})^{2}}{\sigma^{2}_{{{\rm
        sub}}}}\right\}d\delta_{M_{1}} \nonumber\\
&=&\frac{1}{2}\label{eqn:jed}, 
\end{eqnarray}   
where $\sigma^{2}_{{\rm
sub}}=\sigma^{2}_{0}(M_{1})-\sigma^{2}_{0}(M_{2})$ by using the sharp
{\it k}-space filter.  Thus we can obtain the PS formula, naturally
including the factor of two as can be seen from eqs.(\ref{eqn:jed0}) and
(\ref{eqn:jed}).  The reason originating the factor of two is a behavior
of the random walk, and is essentially the same as the excursion set
formalism.
   
However, it is insufficient to use eq.(\ref{eqn:jed}) for more realistic
estimation of the mass function because it is necessary to consider the
spatial correlation of the density fluctuations due to the finite size
of objects.  Besides the sharp $k$-space filter is used which leads
$\sigma_{\rm sub}^{2}$ to $\sigma^{2}_{0}(M_{1})-\sigma^{2}_{0}(M_{2})$,
then we should evaluate how the mass function is changed by other
filters.  Therefore, we must consider the probability $P(r,M_{1}\vert
M_{2})$ of finding $\delta_{M_{1}}\geq\delta_{c}$ at a distance $r$ from
the center of an isolated object of mass scale $M_{2}$ with general
filters.  Then we can get the probability $P(M_1\vert M_2)$ by spatially
averaging $P(r,M_{1}\vert M_{2})$.
   
Because we expect that the isolated collapsed objects are formed
around density peaks, the constraints to obtain the above probability
$P(r,M_{1}\vert M_{2})$ are given as follows:
\begin{enumerate}   
\item The linear density contrast, $\delta_{M_{2}}$, of the larger
mass scale $M_{2}$ should be equal to $\delta_{c}=1.69$ at the center
of the object.
\item Each object of the mass scale $M_{2}$ must contain a maximum peak
of the density field, {\it i.e.}, the first derivative of the density
contrast $\nabla\delta_{M_{2}}$ must be equal to $0$ and each diagonal
component of the diagonalized Hessian matrix $\bfzeta$ of the second
derivatives must be less than $0$ at the center of the object.
\item The density contrast of the smaller mass scale $M_{1}(\leq  
M_{2})$ which has been already collapsed and which is included in an
      object of mass scale $M_{2}$ must satisfy the condition
      $\delta_{M_{1}}\geq\delta_{c}$ at distance $r$ from the center of
      the larger object.
\end{enumerate}   
In order to evaluate the above conditions, in this paper, we investigate
the following effects: (1) filters, (2) density peaks, and (3) spatial
correlation.  Some other effects will be discussed in \S 6.

\section{Results}\label{sec:result}
\subsection{Filtering effect}\label{sec:filter}
In general filters except for the sharp $k$-space filter, the
description of the Markovian random walk in the excursion set formalism
cannot be adopted.  PH90 and BCEK calculated the mass function by Monte
Carlo method realizing Fourier modes $\delta_{\bf k}$ in such a
non-Markov case.  On the other hand, we can easily calculate the mass
function by substituting the conditional probability in
eq. (\ref{eqn:jed}) to general cases as a simple approximation.  It is
adequate approximation for the purpose of this paper in order to
understand the various effects mentioned above, at least, qualitatively.

For this purpose, what we do is to estimate the probability,
\begin{equation}
 P(M_{1}\vert M_{2})=
  p(\delta_{M_{1}}\geq\delta_{c}\vert\delta_{M_{2}}=\delta_{c}).
\label{eqn:fil}
\end{equation}
Note that in the case of general filters, we cannot reduce this
probability to a half because of the correlation between the Fourier
modes.  Here we define the normalized density contrasts,
\begin{eqnarray}
\nu_{1}&\equiv&\frac{\delta_{M_{1}}}{\sigma_{r}},\\
\nu_{2}&\equiv&\frac{\delta_{M_{2}}}{\sigma_{0}},
\end{eqnarray}
where the standard deviations $\sigma_{r}$ and $\sigma_{0}$ are the root
mean squares of the variances $\langle\delta_{M_{1}}^{2}\rangle^{1/2}$
and $\langle\delta_{M_{1}}^{2}\rangle^{1/2}$, which are defined in
Appendix A.  Hereafter we refer values concerning $\delta_{M_{1}}$ to
the subscript $r$.

Generally bivariate Gaussian distribution is given by
\begin{equation}
 p(\nu_{1},\nu_{2}){\rm d}\nu_{1}{\rm d}\nu_{2}=
\frac{1}{2\pi\sqrt{1-\epsilon^{2}}}\exp
\left\{-\frac{(\nu_{1}-\epsilon\nu_{2})^{2}}{2(1-\epsilon^{2})}
-\frac{\nu_{2}^{2}}{2}\right\}
{\rm d}\nu_{1}{\rm d}\nu_{2},
\end{equation}
where the correlation coefficient $\epsilon$ is equal to
$\langle\nu_{1}\nu_{2}\rangle=\xi_{0}(0)/\sigma_{r}\sigma_{0}$, which is
defined in Appendix A.  The conditional probability distribution
function is obtained by the Bayes' theorem,
\begin{eqnarray}
\lefteqn{p(\nu_{1}\vert\nu_{2}=\nu_{2c}){\rm d}\nu_{1}=
\frac{p(\nu_{1},\nu_{2c})}{p(\nu_{2c})}d\nu_{1}}\nonumber\\
&=&\frac{1}{\sqrt{2\pi(1-\epsilon^{2})}}
\exp\left\{-\frac{(\nu_{1}-\epsilon\nu_{2c})^{2}}{2(1-\epsilon^{2})}
\right\}{\rm d}\nu_{1},
\end{eqnarray}
where $\nu_{2c}$ is the normalized critical density contrast with the
smoothing scale $M_{2}$, $\delta_{c}/\sigma_{0}$, and $p(\nu)$ is a
one-point Gaussian normal distribution,
\begin{equation}
 p(\nu)d\nu=\frac{1}{\sqrt{2\pi}}\exp\left(-\frac{\nu^{2}}{2}\right)d\nu.
\end{equation}

In the case of the sharp $k$-space filter, the covariance $\xi_{0}(0)$
is reduced to $\sigma_{0}^{2}$, then we obtain
$\epsilon=\sigma_{0}/\sigma_{r}$.  Thus we obtain the
eq.(\ref{eqn:jed}).  However in general cases, we must calculate
$\xi_{0}(r)$ for specified filters.

In Figs.{\ref{fig:corrfil}} and \ref{fig:corrfilth}, we show
$P(M_{1}\vert M_{2})=\int_{\nu_{1c}}^{\infty}
p(\nu_{1}\vert\nu_{2}=\nu_{2c}){\rm d}\nu_{1}$ in the cases of the
Gaussian filter with the spectral index $n=0$, and the top-hat filter
with $n=-2$ because $\epsilon$ accidentally becomes
$\sigma_{0}/\sigma_{r}$ in $n=0$ and then $P(M_{1}\vert M_{2})=1/2$.  It
is evident that the Fourier mode correlation increases the value of
$P(M_{1}\vert M_{2})$ above a half when $M_{1}\ga M_{*}$, in contrast to
the case of the sharp $k$-space filter, in which the probability is a
half for all region with $M_{2}\geq M_{1}$.

\begin{inlinefigure}
\includegraphics[width=8cm]{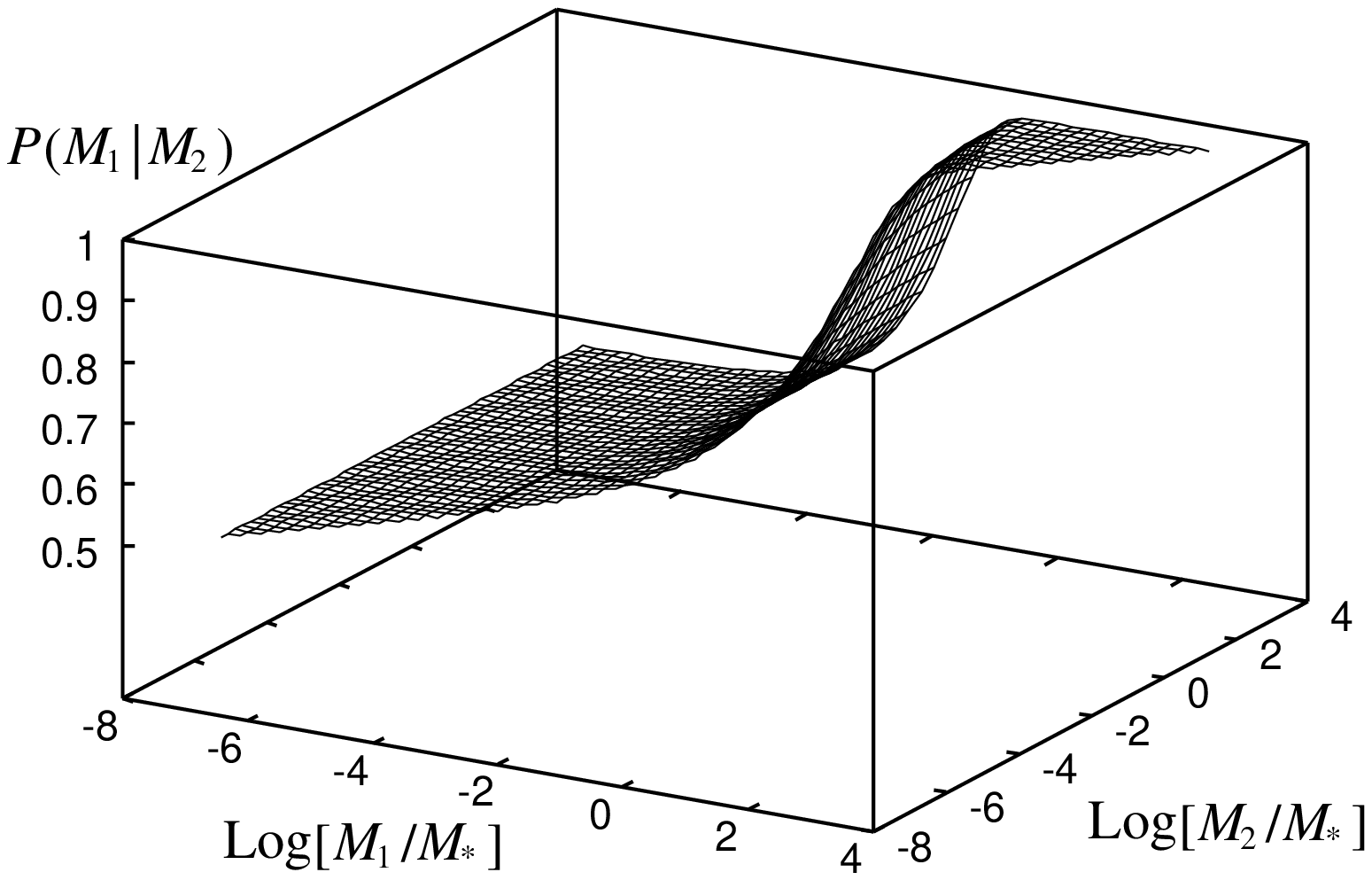}
\caption{Probability $P(M_{1}\vert M_{2})$ for the Gaussian filter in
 the case of the spectral index $n=0$.  The surface of the probability
 is plotted only against the region $M_{1}\leq M_{2}$.}
\label{fig:corrfil}
\end{inlinefigure}

\begin{inlinefigure}
\includegraphics[width=8cm]{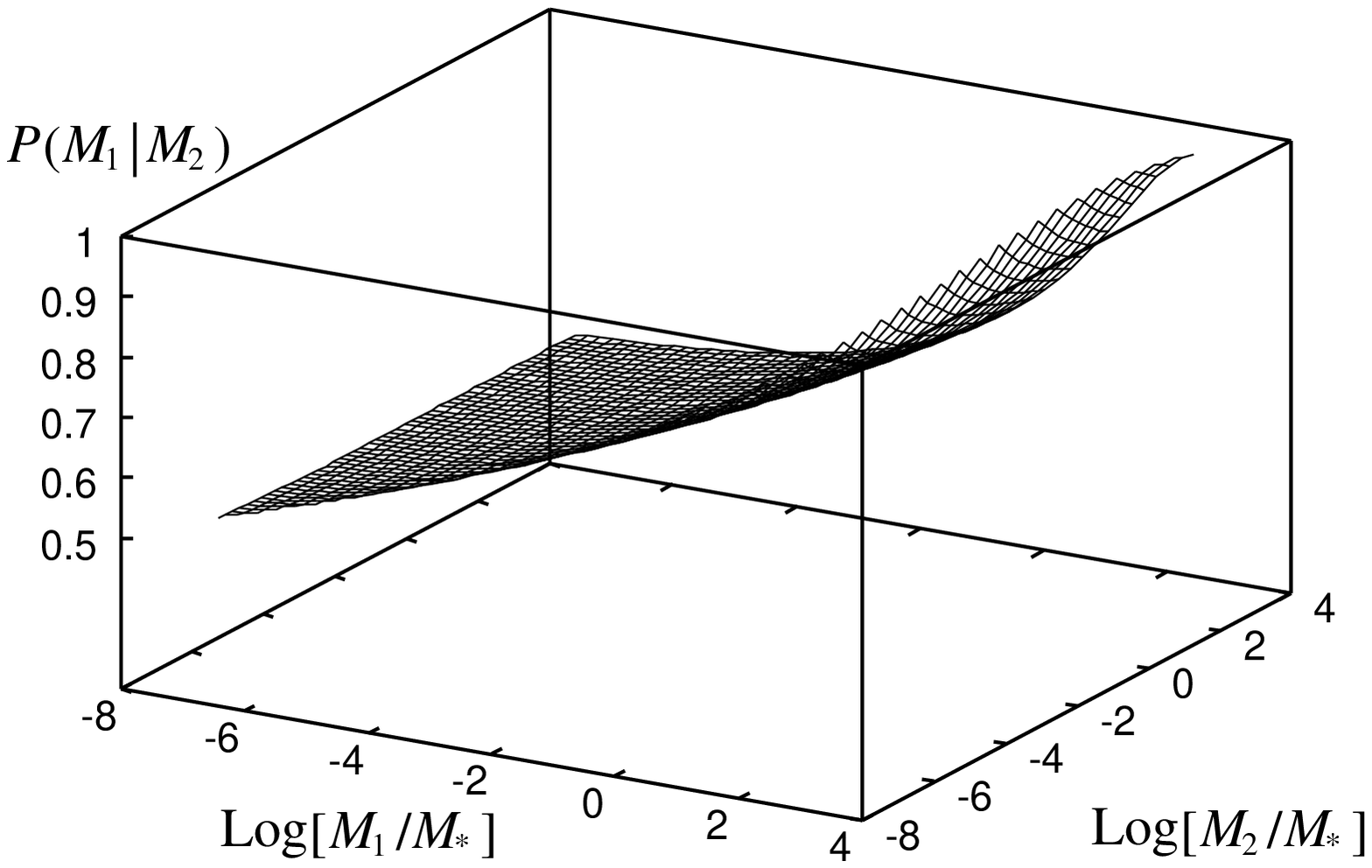} 
\caption{Same as Fig.\ref{fig:corrfil}, but for the top-hat filter and
for $n=-2$.}
\label{fig:corrfilth}
\end{inlinefigure}

Solving eq.(\ref{eqn:jed0}) by substituting $P(M_{1}\vert M_{2})$, we
obtain the mass functions.  In Fig.{\ref{fig:mf1}}, we show the mass
functions for the Gaussian and the top-hat filters and the PS mass
function in a form of a differential multiplicity function, $F(M)\equiv
Mn(M)/\bar{\rho}$.  Note that the function with the sharp $k$-space
filter is the same as the PS mass function.  Because $P(M_{1}\vert
M_{2})$ increases above a half at larger $M_{1}$ scale while
$f(\geq\delta_{c};M_{1})$ is not changed by filters, the number density
decreases at larger scale.  Then, so as to cancel out the decrease, the
number density increases at smaller $M_{2}$ scale.  As a whole, the mass
function is seemed to be moved to lower mass scale.  This result is
similar to that given by the Monte Carlo calculation by PH90 and BCEK.
Thus we conclude that the filtering effect can be interpreted as the
correlation effect between different Fourier modes $\delta_{\bf k}$.  It
should be noted that in the case of the top-hat filter and $n=0$, the
correlation $\xi_{0}(0)$ becomes $\sigma_{0}^{2}$, so the resultant mass
function is exactly the same as the PS mass function.

\begin{inlinefigure}
\includegraphics[width=8cm]{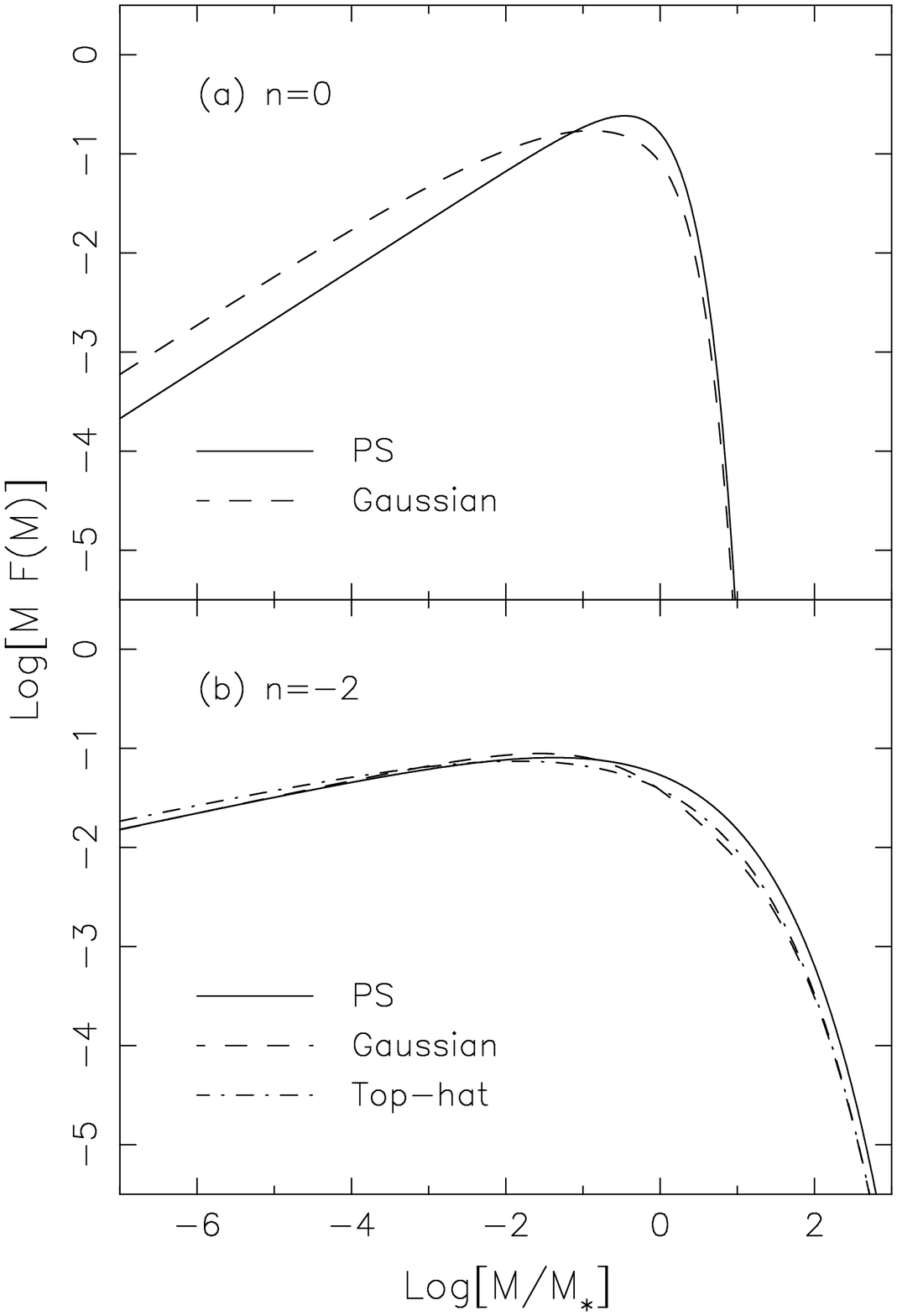}
\caption{Differential multiplicity functions $F(M)$ with the filtering
 effect.  (a) $n=0$.  (b) $n=-2$.  The solid, the dashed, and the
 dash-dotted lines denote $F(M)$ for the PS, the Gaussian filter, and
 the top-hat filter.  Note that $F(M)$ for the sharp $k$-space filter
 and for the top-hat filter with $n=0$ are identical with the PS.}
 \label{fig:mf1}
\end{inlinefigure}

\subsection{Peak effect}\label{sec:peak}
Next we derive the mass function under the assumption that objects
collapse around density maxima.  Because ten variables are required to
describe one peak, we need to calculate eleven-variate Gaussian
distribution function (one is $\delta_{M_{1}}$ and others are concerning
an object $M_{2}$).  Here we define the first and the second derivatives
of the density contrast $\delta$,
\begin{eqnarray}
 \eta_{i}&\equiv&\frac{\partial\delta({\bf x})}{\partial x_{i}},\\
 \zeta_{ij}&\equiv&\frac{\partial^{2}\delta({\bf x})}
{\partial x_{i}\partial x_{j}}.
\end{eqnarray}
The peak condition is described by using $\eta_{i}$ and $\zeta_{ij}$ as
follows: $\bfeta=0$ and the eigenvalues of the Hesse matrix $\bfzeta$
are less than 0.  The covariances concerning $\eta_{i}$ and $\zeta_{ij}$ are estimated as
follows:
\begin{eqnarray}
\langle\eta_{i}\eta_{j}\rangle&=&\frac{\sigma_{1}^{2}}{3}\delta_{ij},\\
\langle x'\nu_{1}\rangle&\equiv&\mu(r=0)=\frac{\xi_{1}(r=0)}
{\sigma_{r}\sigma_{2}},\label{eqn:defmu}\\
\langle x'\nu_{2}\rangle&\equiv&\gamma=\frac{\sigma_{1}^{2}}{\sigma_{0}\sigma_{2}}\label{eqn:defgamma},
\end{eqnarray}
where $x'$ is a normalized trace of the diagonalized Hesse matrix,
\begin{equation}
 x'\equiv\frac{\lambda_{1}+\lambda_{2}+\lambda_{3}}{\sigma_{2}},
\end{equation}
and the variances and correlation are defined in Appendix A.  Note that
$(-\lambda_{1}, -\lambda_{2}, -\lambda_{3})$ are the eigenvalues of
$\bfzeta$, and other three components are transformed to variables given
by the Euler angles for the diagonalization.

By using the above quantities, we can calculate the probability
$P(M_{1}\vert M_{2})$ as follows,
\begin{eqnarray}   
P(M_{1}\vert M_{2})&=&
 p(\nu_{1}\geq\nu_{c}\vert\nu_{2}=\nu_{2c},\bfeta=0,\lambda_{i=1,2,3}>0)\nonumber\\
 &=&\sqrt{\frac{1-\gamma^{2}}{2\pi(1-\epsilon^2-\mu^{2}-\gamma^{2}+
  2\epsilon\mu\gamma)}}\nonumber\\
 &&\times\frac{\int_{0}^{\infty}dx'~f(x')\int_{\nu_{1c}}
  ^{\infty}d\nu_{1}\exp(-\frac{Q_{a}+Q_{b}}{2})}{\int_{0}^{\infty}
  dx'~f(x')\exp(-\frac{Q_{b}}{2})},
\label{eqn:peak}
\end{eqnarray}   
where $Q_{a}, Q_{b}$ and $f(x)$ are defined in Appendix A.  Substituting
the above probability to eq.(\ref{eqn:jed}), we obtain mass functions.

It should be noted that in the case of the top-hat filter, the second
derivative cannot be done because the top-hat filter does {\it not}
smooth out the density field adequately.  Then the variance for the
derivatives of the density contrast diverges to infinity as shown in
Appendix B.  This fact leads that the resultant mass function taking
into account the peak condition becomes the same as that not taking into
account it, if we simply interpret this divergence as $\gamma, \mu\to 0$
in eqs.(\ref{eqn:defmu}) and (\ref{eqn:defgamma}).  The peak formalism
should be modified to be able to treat the top-hat filter but it is
beyond the scope of this paper.

Apparently the similar thing to the above occurs for the sharp $k$-space
filter.  If we do not consider the spatial correlation of the density
field, the peak condition does not add any new properties because the
behavior of the Markovian random walk owing to adding a new Fourier mode
conserves.  Mathematically this is proved by the fact that
$\mu=\epsilon\gamma$ in the case of the sharp $k$-space filter.  Thus we
obtain the same mass function as the PS even taking into account the
peak condition.  Of course, when we consider the spatial correlation,
the peak condition will affect the resultant mass function because $\mu$
does not coincide with $\epsilon\gamma$ at $r\neq 0$ (see the next
subsection).

In Fig.{\ref{fig:corrpk}}, we show $P(M_{1}\vert M_{2})$ in the case of
the Gaussian filter with the spectral index $n=0$.  Note that the
viewing angle is changed.  We see that $P(M_{1}\vert M_{2})$ shows a
behavior such as a function of $(M_{1}/M_{2})$.  Thus the peak condition
with the Fourier mode correlation increases the value of $P(M_{1}\vert
M_{2})$ above a half at $M_{2}\sim M_{1}$ even at $M_{1}\ll M_{*}$, in
contrast to the results in the previous subsection.

\begin{inlinefigure}
\includegraphics[width=8cm]{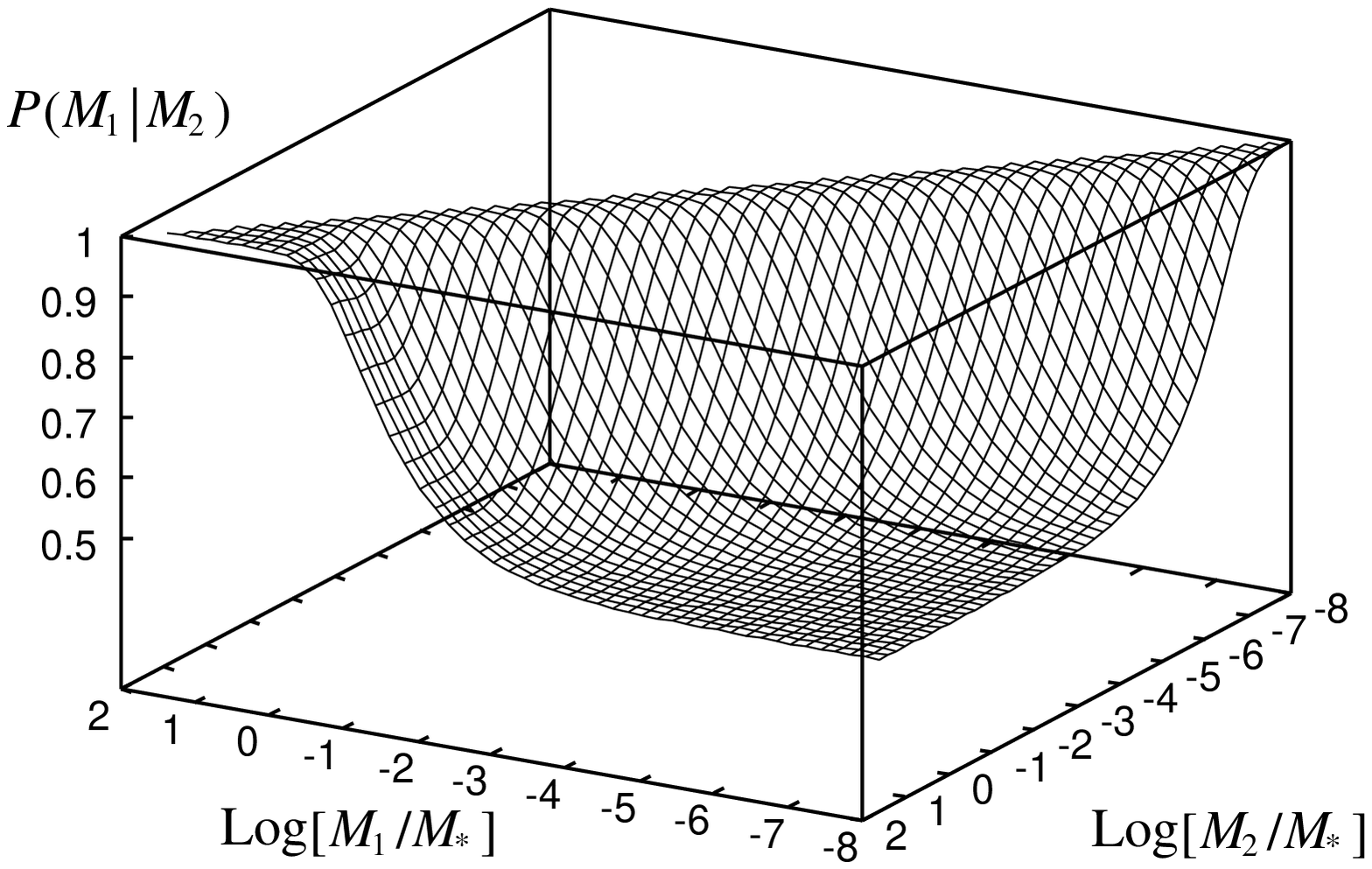}
\caption{Probability $P(M_{1}\vert M_{2})$ with the peak effect for the
 Gaussian filter in the case of $n=0$.  The viewing angle is changed in
 order to easily see the shape of the probability from
 Figs.\ref{fig:corrfil} and \ref{fig:corrfilth}.}
\label{fig:corrpk}
\end{inlinefigure}

In Fig.\ref{fig:mf2}, we show the resultant mass functions for the
Gaussian filter.  For the references, we also show the mass functions of
the PS and the top-hat filter derived in the previous subsection.  In
the case of $n=0$, the Gaussian filter affects the mass function
especially at lower mass scale, $M\la M_{*}$ and the slope at that scale
becomes shallower than that of the PS.  This change reflects the shape
of $P(M_{1}\vert M_{2})$ near $M_{1}\sim M_{2}$.  In the case of $n=-2$,
similar to the case of $n=0$, the slope at lower scale is changed.
However the difference from the PS decreases compared with the $n=0$
case.  This will lead that the filtering and the peak effects become
small in the realistic CDM power spectrum with $n\la -2$ at galactic
scale.  Note that PH90 also obtained a similar mass function taking into
account the peak condition with another approximation scheme.

\begin{inlinefigure}
\includegraphics[width=8cm]{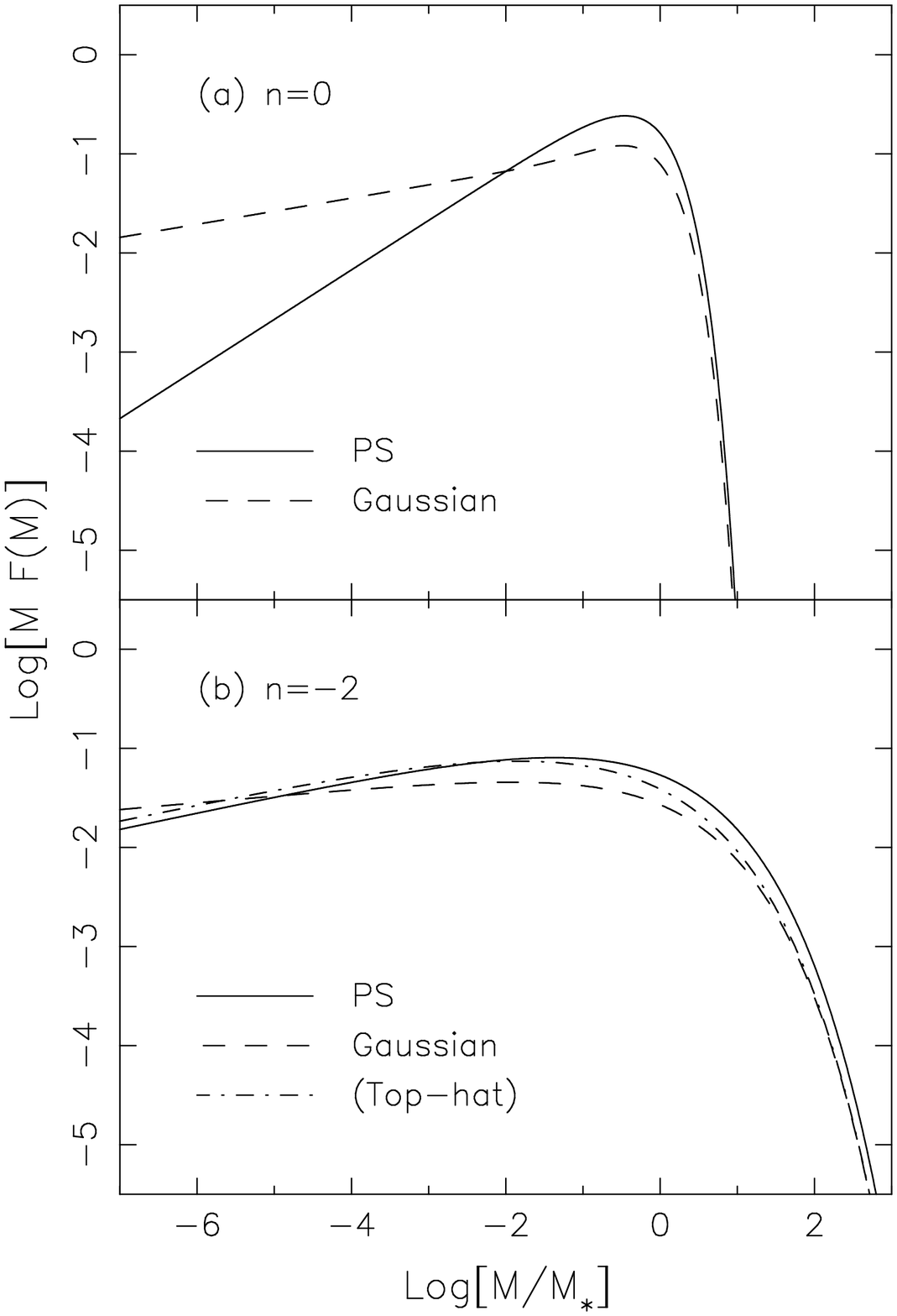}
\caption{Differential multiplicity functions $F(M)$ with the peak
 effect.  (a) $n=0$.  (b) $n=-2$.  The solid and the dashed lines denote
 $F(M)$ for the PS and the Gaussian filter.  For the reference, the
 function for the top-hat filter only with the filtering effect is shown
 by the dash-dotted line because the peak effect cannot be dealt
 correctly.  As mentioned in Fig.\ref{fig:mf1}, $F(M)$ for the sharp
 $k$-space filter and for the top-hat filter with $n=0$ are identical
 with the PS.}
\label{fig:mf2}
\end{inlinefigure}

\subsection{spatial correlation}\label{sec:spcorr}
Finally, we evaluate the finite-size effect, that is, the effect of the
spatial correlation on the mass function.  As already shown in YNG, this
effect is similar to changing the critical density $\delta_{c}$ in the
PS to a smaller value than 1.69, in the case of the sharp $k$-space
filter.  Here we generalize the YNG result to general filters.

First we consider only the spatial correlation without the peak effect.
In this case, the correlation coefficient $\epsilon$ must be generalized
to that between two distant points with different mass scales, that is,
$\epsilon\to\epsilon(r)$, and the probability eq.(\ref{eqn:fil}) is
interpreted as $P(r,M_{1}\vert M_{2})$.  Then by spatial averaging over
the region of the object $M_{2}$, we obtain the probability
$P(M_{1}\vert M_{2})$,
\begin{equation}   
P(M_{1}\vert M_{2})=\frac{\int_{0}^{R_{2}}P(r,M_{1}\vert M_{2})   
        4\pi r^{2}dr}{\int_{0}^{R_{2}}4\pi r^{2}dr}.\label{eqn:ave}
\end{equation}   
It should be noted that it is not trivial that the window function
$W_{M_{2}}(r)$ is not needed in this averaging.  However, since the
averaging operation has been already executed in the calculations of
$\delta_{M_{1}}$ and $\delta_{M_{2}}$, we estimate $P(M_{1}\vert M_{2})$
by eq.(\ref{eqn:ave}).  When it is included, the effect of the spatial
correlation will be enhanced because more distant point from the center
of the object $M_{2}$ is considered.

\begin{inlinefigure}
\includegraphics[width=8cm]{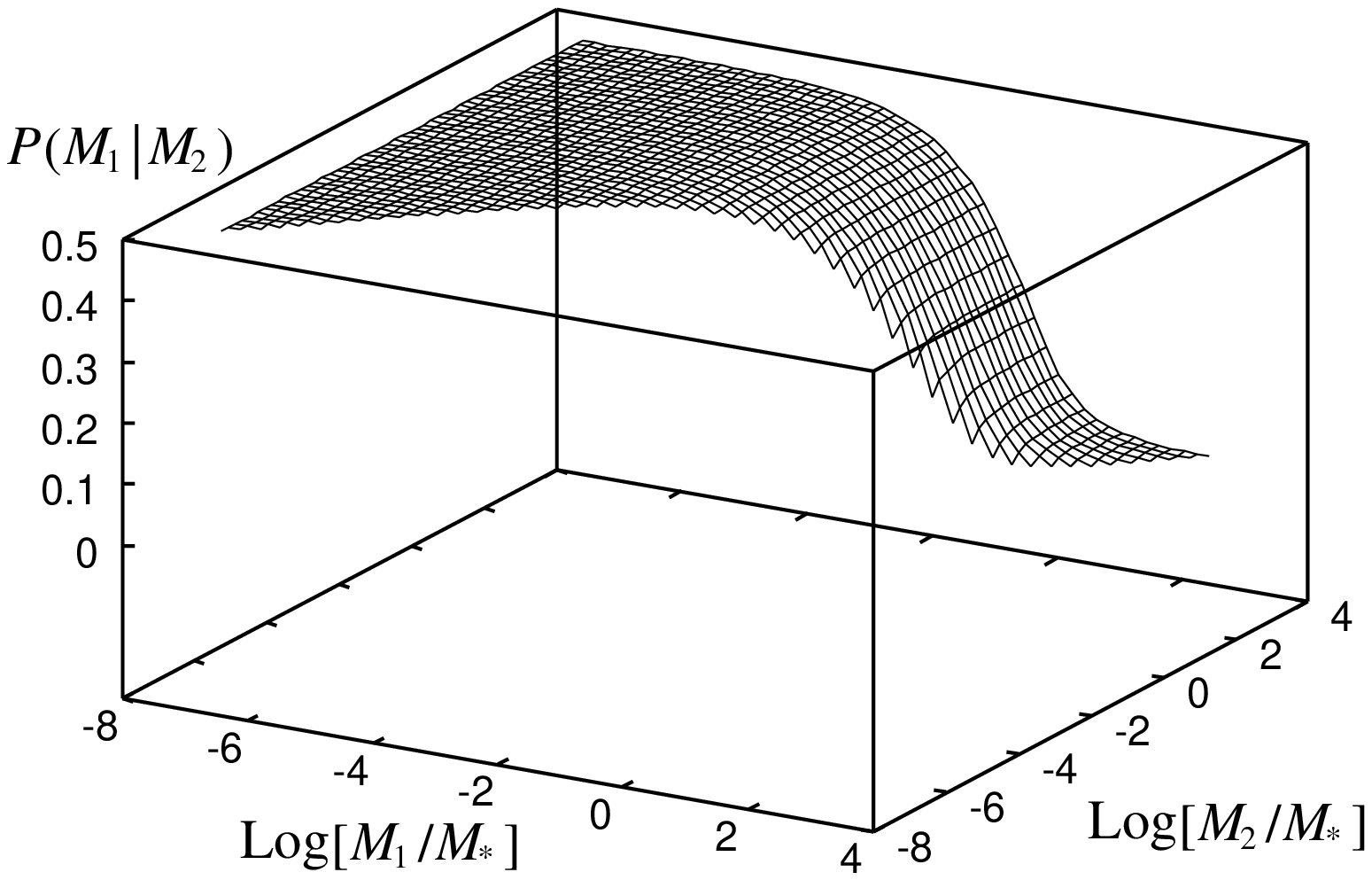}
\caption{Probability $P(M_{1}\vert M_{2})$ with the effect of the
 spatial correlation for the sharp $k$-space filter in the case of
 $n=0$.}
\label{fig:corrsp1}
\end{inlinefigure}

\begin{inlinefigure}
\includegraphics[width=8cm]{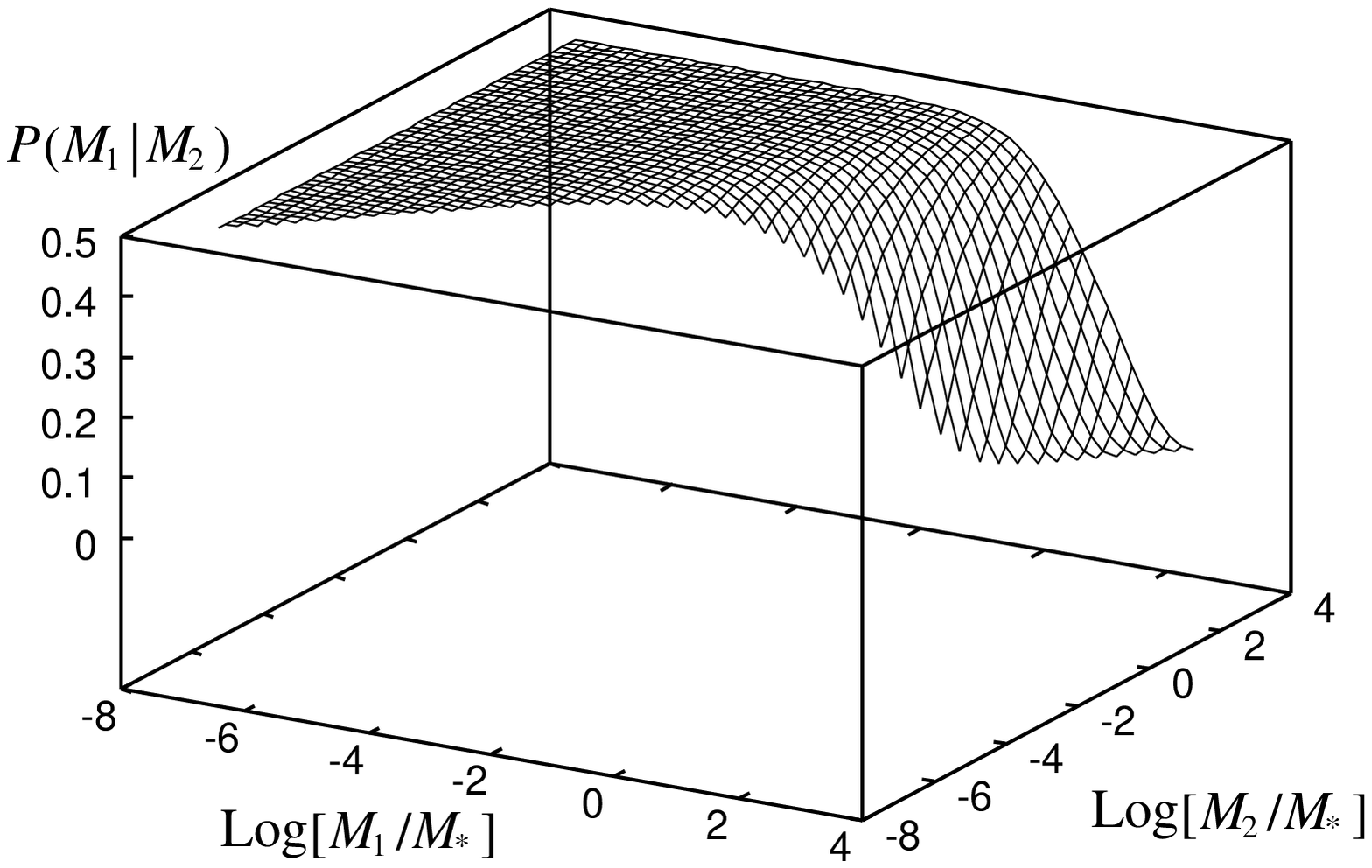}
\caption{Same as Fig.\ref{fig:corrsp1}, but for the top-hat filter.}
\label{fig:corrsp2}
\end{inlinefigure}

\begin{inlinefigure}
\includegraphics[width=8cm]{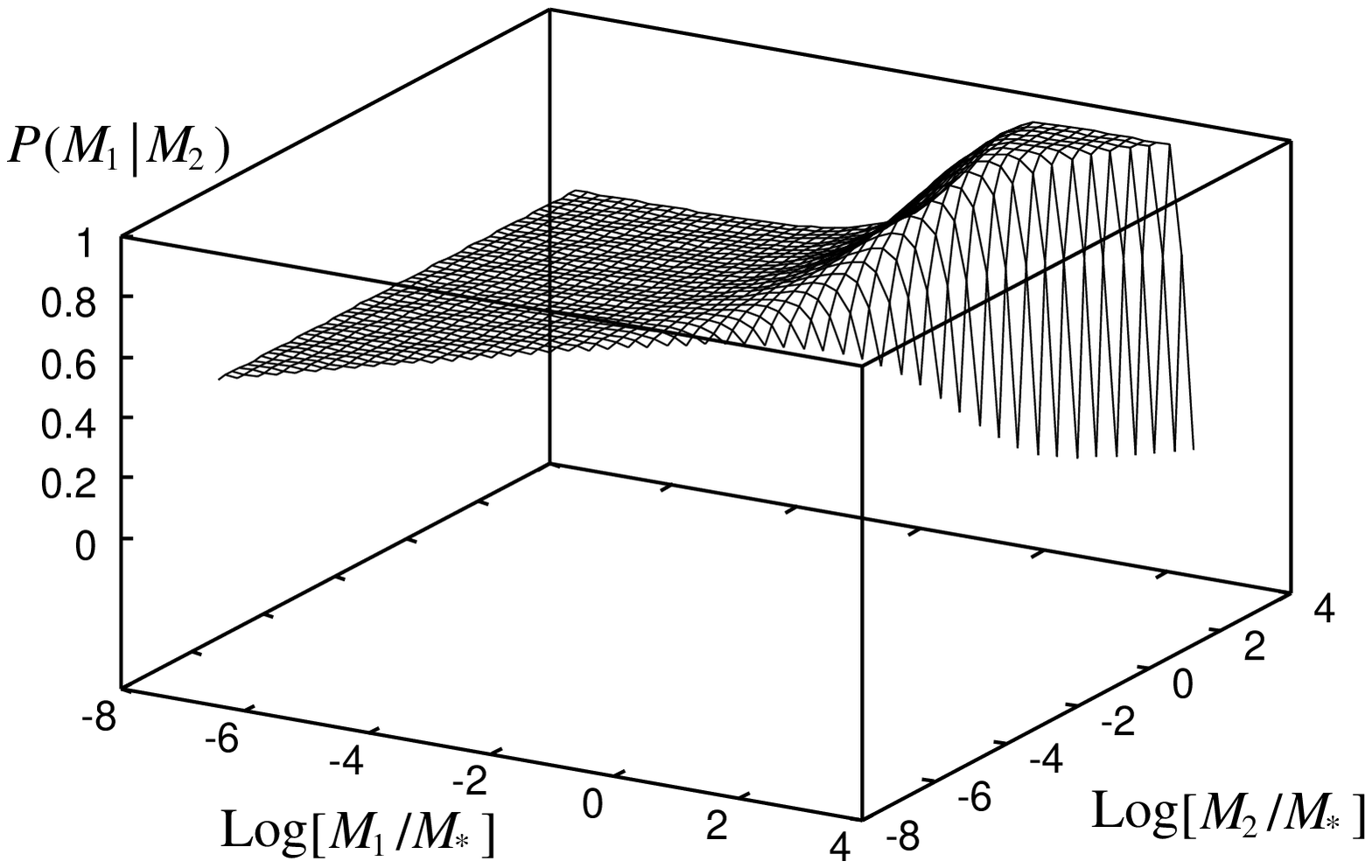}
\caption{Same as Fig.\ref{fig:corrsp1}, but for the Gaussian filter.}
\label{fig:corrsp3}
\end{inlinefigure}

In Figs.\ref{fig:corrsp1} $\sim$\ref{fig:corrsp3}, we show the spatially
averaged probabilities $P(M_{1}\vert M_{2})$ for the sharp $k$-space,
the top-hat, and the Gaussian filters with the spectral index $n=0$.  In
the cases of the two former filters, in contrast to the filtering
effect, the probability becomes less than a half when $M_{1}\ga M_{*}$.
The reason why the value becomes less than a half is that the
correlation between $\delta_{M_{1}}$ and $\delta_{M_{2}}$ becomes weaker
at more distant point from the center of the object $M_{2}$.  Thus,
while the interval of the integration of $\delta_{M_{1}}-\delta_{c}$ is
[$0,\infty$] and then the probability is a half when the sharp $k$-space
filter is used and the spatial correlation is neglected
[eq.(\ref{eqn:jed})], we must integrate from larger than 0 to $\infty$
under the consideration of the finite volume.  Then the probability
becomes less than a half.  On the other hand, in the Gaussian filter,
the decrease of $P(M_{1}\vert M_{2})$ at $M_{1}\ga M_{*}$ emerges only
at $M_{1}\simeq M_{2}$.  Except for such region $M_{1}\simeq M_{2}\ga
M_{*}$, the shape of $P(M_{1}\vert M_{2})$ is similar to
Fig.\ref{fig:corrfil} in which the spatial correlation is not taken into
account.  This suggests that the filtering effect is stronger than the
effect of the spatial correlation in the Gaussian filter.

We show the mass functions for three filters in Fig.\ref{fig:mf3}.
While the mass functions with the top-hat and the sharp $k$-space
filters horizontally move to larger mass scale compared with the PS mass
function, the mass function with the Gaussian filter moves to smaller
mass scale.  These properties can be explained from the shape of
$P(M_{1}\vert M_{2})$.  In the sharp $k$-space and the top-hat filters,
$P(M_{1}\vert M_{2})$ is less than a half at $M_{1}\ga M_{*}$.  This
leads that the multiplicity function $F(M)$ becomes larger than the PS
at that scale because $f(\geq\delta_{c};M_{1})$ in eq.(\ref{eqn:jed0})
is unchanged.  Then $F(M)$ decreases at smaller scale $M_{1}\la M_{*}$
so as to cancel the increase of $F(M)$ at the larger scale.  On the
contrary, in the Gaussian filter, $P(M_{1}\vert M_{2})$ increases above
a half in almost the region of $M_{1}\ga M_{*}$.  This leads that $F(M)$
becomes smaller at that scale and larger at smaller scale.

\begin{inlinefigure}
\includegraphics[width=8cm]{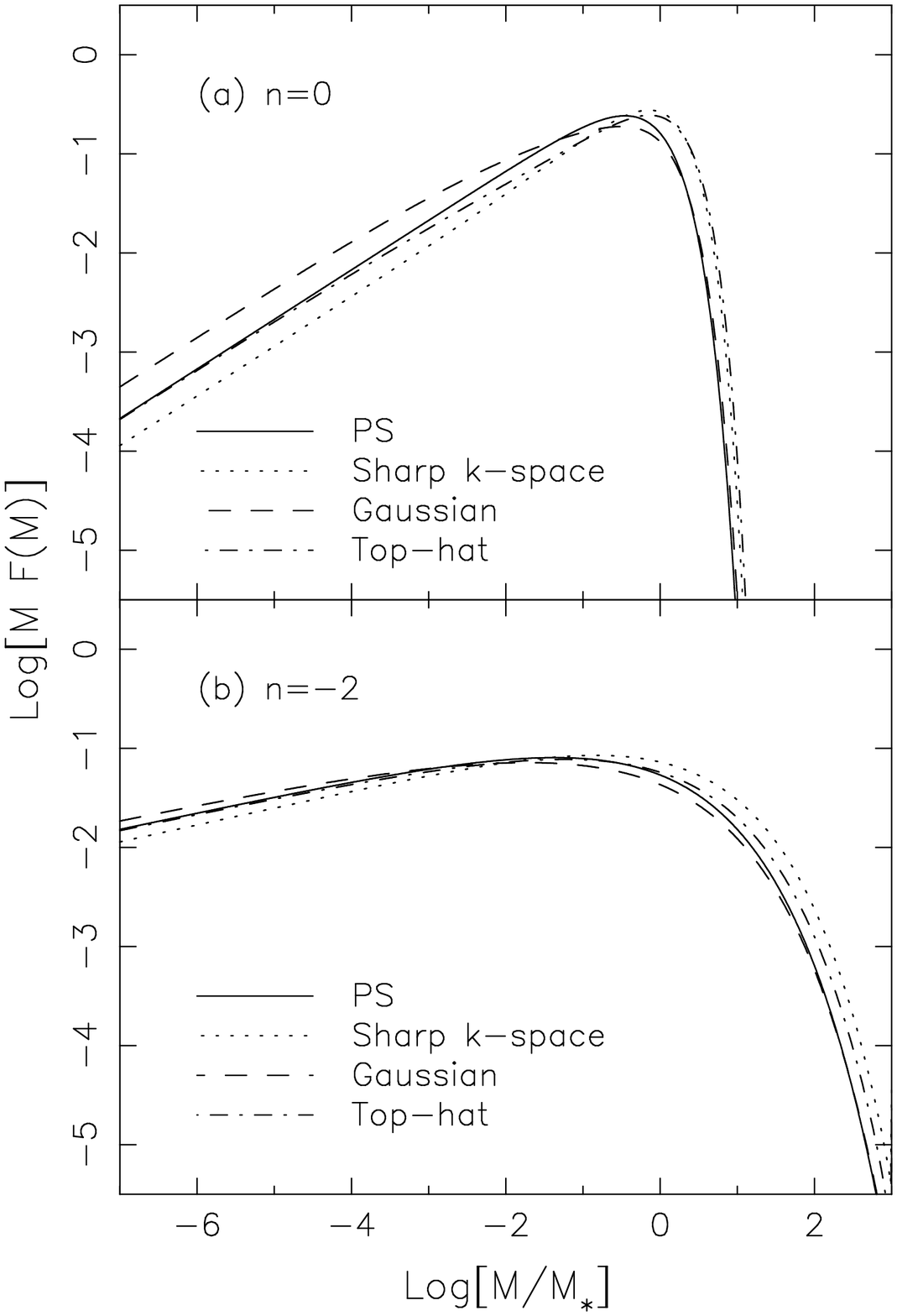}
\caption{Differential multiplicity functions $F(M)$ with the effect of
 the spatial correlation.  (a) $n=0$.  (b) $n=-2$.  The solid, the
 dotted, the dashed and the dash-dotted lines denote $F(M)$ for the PS,
 the sharp $k$-space filter, the Gaussian filter and the top-hat filter.}
\label{fig:mf3}
\end{inlinefigure}

Finally we include the peak condition in the above formalism.  In this
case, not only $\epsilon(r)$ but also $\mu$ must be generalized to
$\mu(r)$ because $\mu(r)$ describes the correlation between the second
derivative of the density maxima with smoothing scale $M_{2}$ and the
density height with scale $M_{1}$ distant $r$ from the center of the
object $M_{2}$.  In order to obtain the probability $P(M_{1}\vert
M_{2})$, we substitute $\epsilon(r)$ and $\mu(r)$ into
eq.(\ref{eqn:peak}) and average over the region of the object $M_{2}$
according to eq.(\ref{eqn:ave}).

\begin{inlinefigure}
\includegraphics[width=8cm]{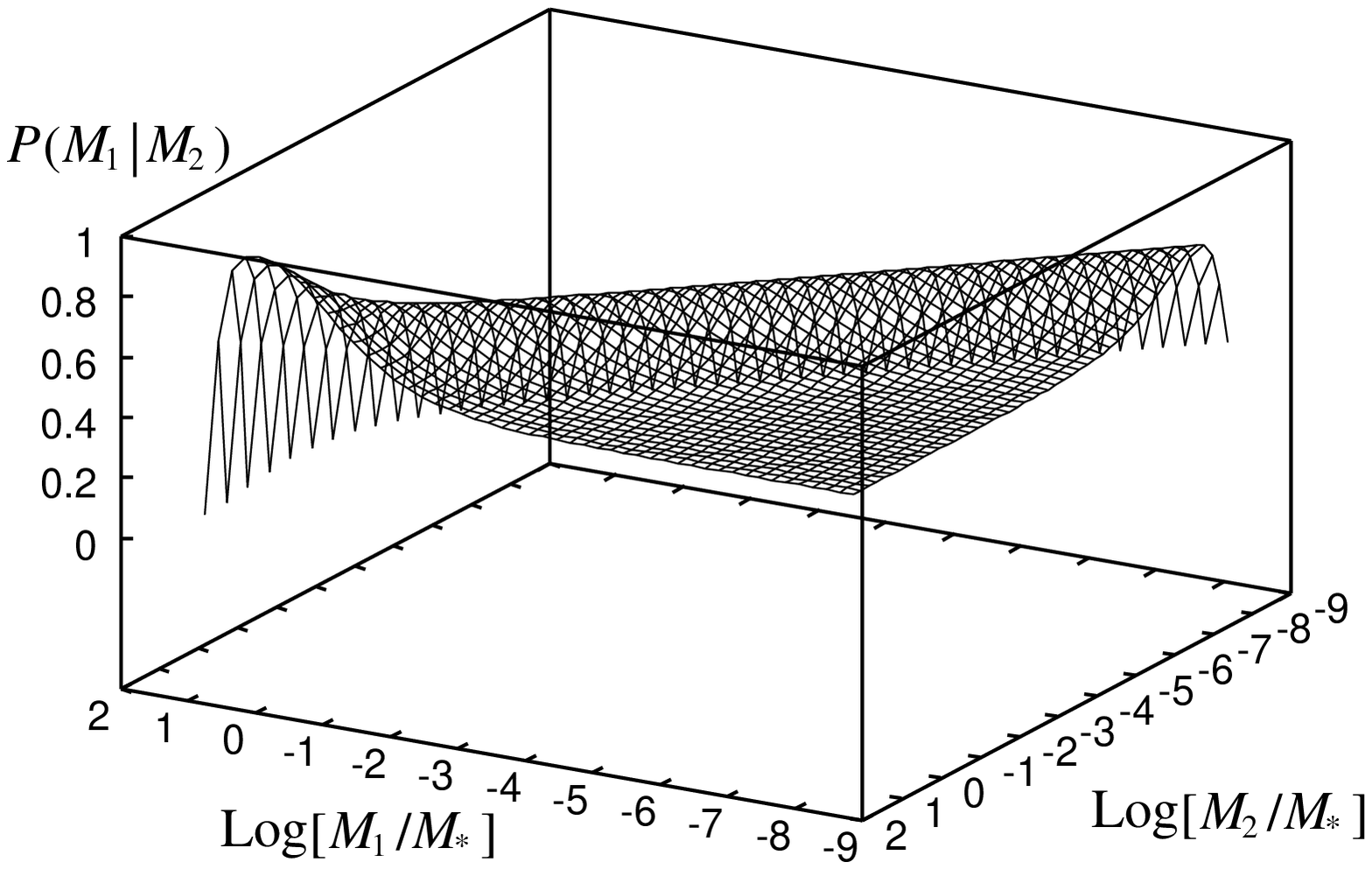}
\caption{Probability $P(M_{1}\vert M_{2})$ with the effects of the peak
 and the spatial correlation for the Gaussian filter in the case of
 $n=0$.}
\label{fig:corrpc1}
\end{inlinefigure}

\begin{inlinefigure}
\includegraphics[width=8cm]{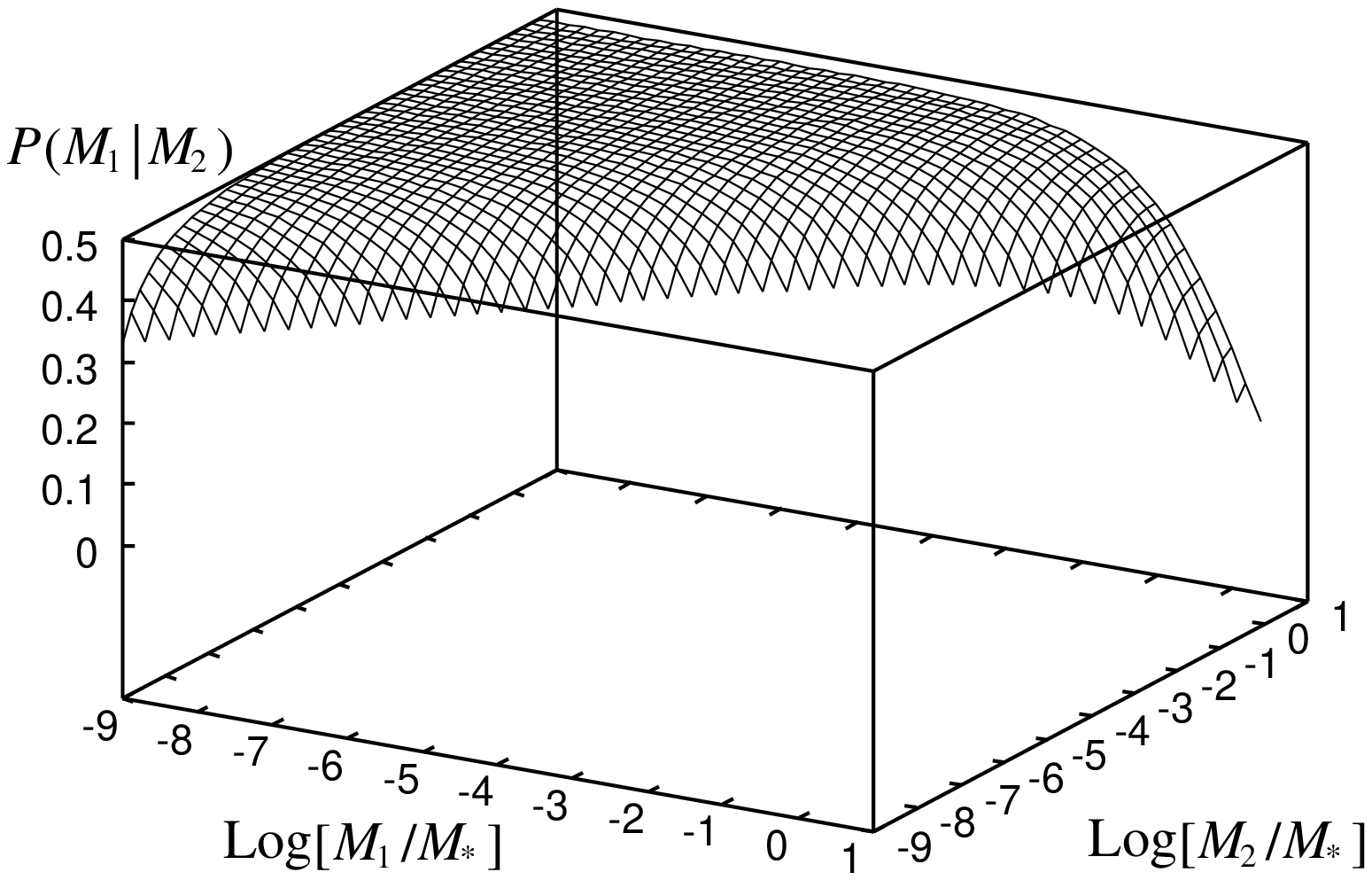}
\caption{Same as Fig.\ref{fig:corrpc1}, but for the sharp $k$-space
filter.}
\label{fig:corrpc2}
\end{inlinefigure}

In Fig.\ref{fig:corrpc1}, we show $P(M_{1}\vert M_{2})$ for the Gaussian
filter with the peak and spatial correlation effects.  This figure shows
the following three characteristics: increase at $M_{2}\gg M_{1}\ga
M_{*}$, increase toward to $M_{2}\sim M_{1}$, and steep decrease at
$M_{2}\simeq M_{1}$.  It can be easily interpreted from
Figs.\ref{fig:corrfil} and \ref{fig:corrpk} that the first
characteristic reflects the filtering effect and the second the peak
effect.  On the other hand, the third is a new property generated by the
mixture of the filtering, peak and spatial correlation effects, because
the spatial correlation itself affects $P(M_{1}\vert M_{2})$ only at
$M_{1}\ga M_{*}$ as shown in Fig.\ref{fig:corrsp3}.
Fig.\ref{fig:corrpc2} shows $P(M_{1}\vert M_{2})$ for the sharp
$k$-space filter.  Similar to the third effect in the Gaussian filter
case, $P(M_{1}\vert M_{2})$ decreases at $M_{2}\simeq M_{1}$.  This is
also a new characteristic.  As seen in the peak effect
(Fig.\ref{fig:corrpk}), the change of $P(M_{1}\vert M_{2})$ from a half
at $M_{2}\simeq M_{1}$ affects the slope of the low-mass tail of the
multiplicity function.

In Fig.\ref{fig:mf4}, we show the mass functions taking into account the
effects of filtering, peak, and spatial correlation.  For the reference,
we also show the mass functions for the top-hat filter with the effect
of the spatial correlation (Fig.\ref{fig:mf3}), because we cannot derive
those with the peak effect as mentioned in \S\S\ref{sec:peak}.  At
larger scale $M_{1}\ga M_{*}$, the effect of the spatial correlation is
similar to that without the peak effect shown in Fig.\ref{fig:mf3}.  At
smaller scale $M_{1}\la M_{*}$, as shown by YNG, the slope becomes
steeper than that of the PS in the case of the sharp $k$-space filter.
In contrast, in the case of the Gaussian filter, the slope becomes much
shallower than the PS.  The effect of the spatial correlation is too
weak to cancel out the peak effect which makes such a shallow slope,
while at larger scale the mass function is close to the PS.

These results show that the slope at smaller scale $M_{1}\ll M_{*}$ is
determined by the behavior of $P(M_{1}\vert M_{2})$ from the region
$M_{1}\ll M_{2}$ at which $P(M_{1}\vert M_{2})$ is a half toward
$M_{1}\sim M_{2}$.  If $P(M_{1}\vert M_{2})$ has a region $P(M_{1}\vert
M_{2})>1/2$, the slope becomes shallow and if $P(M_{1}\vert M_{2})<1/2$
near $M_{1}\sim M_{2}$, the slope becomes steep.  In the case of the
top-hat filter, in which the peak effect cannot be considered,
$P(M_{1}\vert M_{2})\simeq 1/2$ at $M_{2}\sim M_{1}\la M_{*}$, then the
slope is close to that of the PS.  Note that the difference of the slope
from the PS is prominent in large spectral index, $n$.  We will comment
on the power spectrum in later section.

\begin{inlinefigure}
\includegraphics[width=8cm]{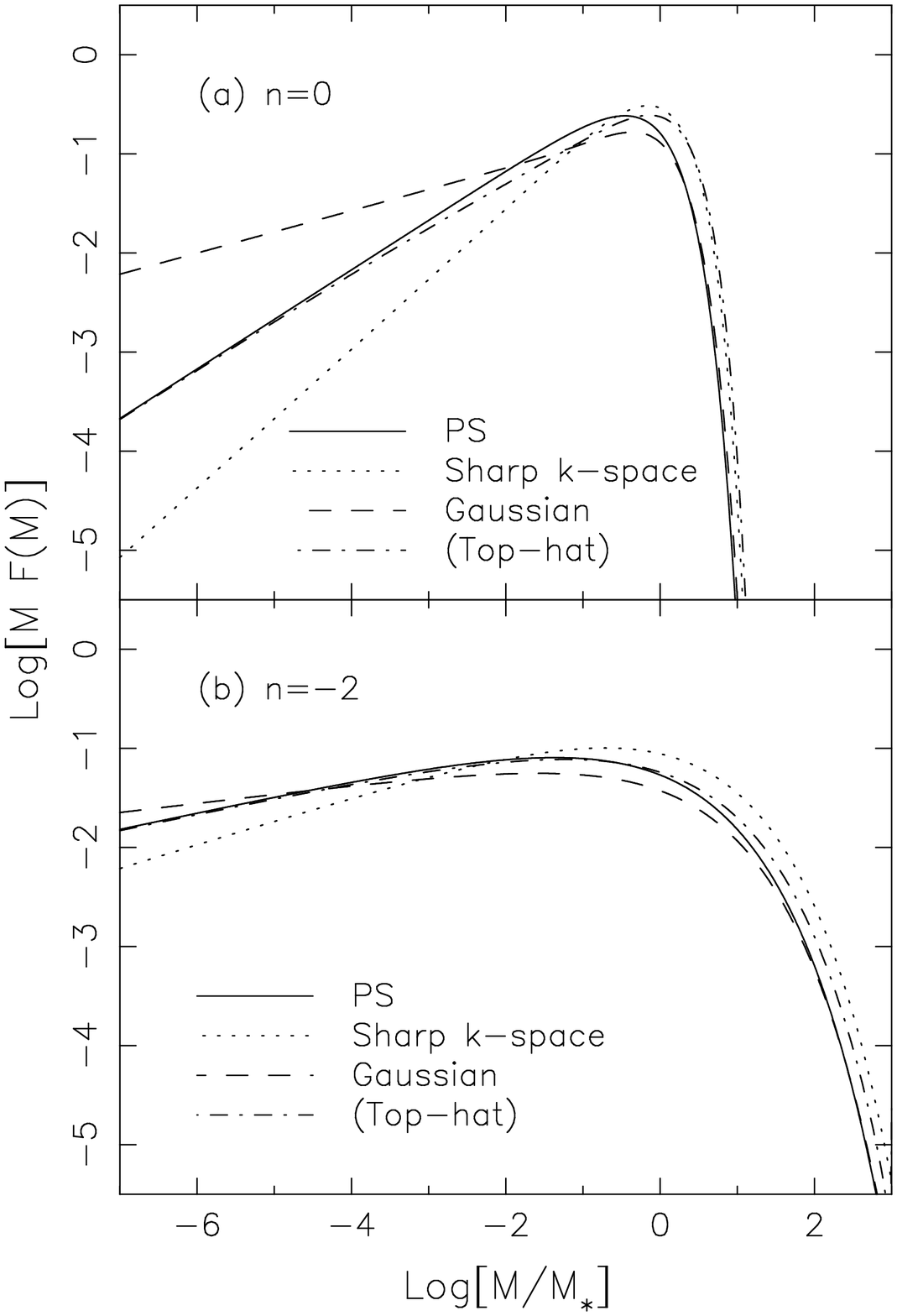}
\caption{Differential multiplicity functions $F(M)$ with the effects of
 the peak and the spatial correlation.  (a) $n=0$.  (b) $n=-2$.  The
 solid, the dotted and the dashed lines denote $F(M)$ for the PS, the
 sharp $k$-space filter and the Gaussian filter.  For the reference the
 functions for the top-hat filter only with the effect of the spatial
 correlation are shown by the dash-dotted lines.}
\label{fig:mf4}
\end{inlinefigure}

\subsection{CDM power spectrum}\label{sec:CDM}

In this subsection we show the mass function in the case of a flat
$\Lambda$CDM model as a example of realistic cases.  We use a power
spectrum given by BBKS with $\Omega_{0}=0.3, \Omega_{\Lambda}=0.7$ and
$\Gamma=0.21$, where $\Omega_{0}, \Omega_{\Lambda}$ and $\Gamma$ are the
cosmological density parameter, the cosmological constant and the shape
parameter, respectively.  Here we assign the smoothing scale $R_{\rm G}$
for the Gaussian filter to 0.64$R$ according to BBKS and define the mass
of objects $M=6\pi^{2}\bar{\rho}k_{c}^{-3}$ according to LC for the
sharp $k$-space filter.  The power spectrum is normalized so as to be
unity at $8h^{-1}$Mpc with the top-hat filter.  In Fig.\ref{fig:mf5}, we
show the mass functions for the three filter functions with the effect
of the spatial correlation and the PS mass function.  We neglect the
peak effect in order to compare the mass functions under the same
condition.

The behavior for filter changing is similar to the scale-free power
spectra (see Fig.\ref{fig:mf3}) and the difference among them becomes a
little larger.  Note that there is an uncertainty in the estimation of
mass of objects in the cases of the Gaussian and the sharp $k$-space
filters and it cannot be canceled in the case of such realistic power
spectrum with a characteristic scale.  If we adopt the same definition
of mass as in the previous, the difference becomes still larger.

We also plot the mass function given by Jenkins et al. (2001), which is
well fitted by their high resolution $N$-body simulation.  At larger
mass scale $M\ga 10^{15}M_{\odot}$, it is in good agreement with the
mass function with the top-hat filter.  This agreement suggests that the
spherical collapse approximation is good in such large scale ($M\ga
10^{15}M_{\odot}$) and that an overlapping effect between similar mass
halos becomes important in the intermediate scale ($M\sim
10^{14}M_{\odot}$) as shown by YNG.

\begin{inlinefigure}
\includegraphics[width=8cm]{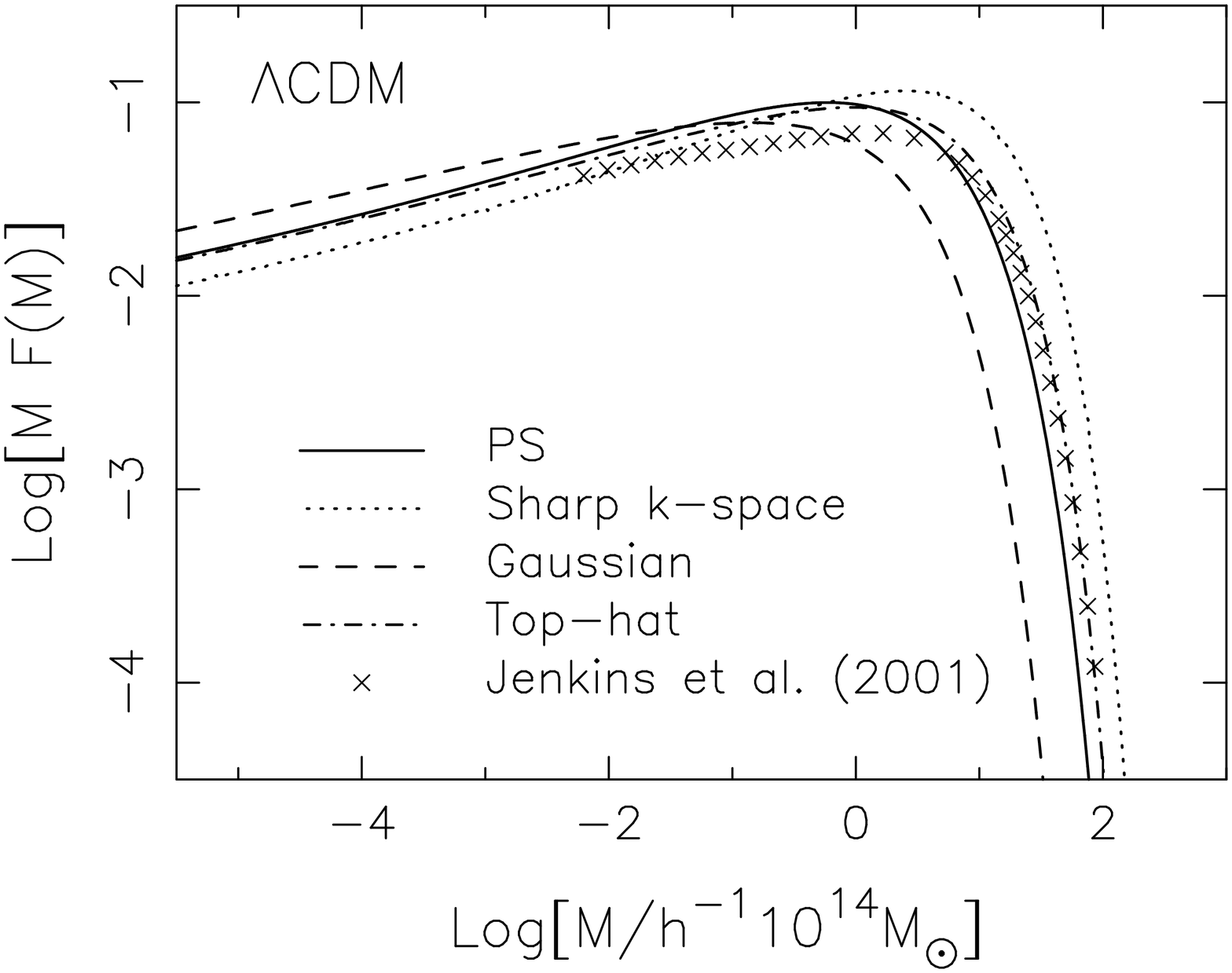}

\caption{Differential multiplicity functions $F(M)$ with the effect of
 the spatial correlation in a $\Lambda$CDM model.  The solid, the
 dotted, the dashed and the dash-dotted lines denote $F(M)$ for the PS,
 the sharp $k$-space, the Gaussian and the top-hat filters.  The crosses
 indicate the mass function given by Jenkins et al. (2001), which was
 obtained by fitting their $N$-body simulation. }

\label{fig:mf5}
\end{inlinefigure}

\section{Discussion}
In this section we discuss some uncertainties in calculation of the mass
function.

\subsection{Mass-smoothing scale relation}
So far we use a simple mass-smoothing scale relation, $M=4\pi
R^{3}\bar{\rho}/3$ and $k_{c}R=\pi$ in the case of scale-free power
spectra.  By changing this relation, the mass function may be affected
as follows.  It should be noted that we calculate the mass function
against the normalized mass $M/M_{*}$ and $M_{*}$ is defined so as to
$\sigma_{0}(M_{*})=1$ with scale-free power spectra.  Thus the mass
function does not scale in mass by changing the relation, even in the
case of the Gaussian filter.  This effect appears via the spatial
correlation in the case of the sharp $k$-space filter.

The ratio of the cut-off wave number for the simple relation ($k_{c}$)
to that given by the LC method ($k_{c,\rm LC}$), $k_{c}/k_{c, \rm
LC}=(2\pi^{2}/9)^{1/3}\sim 1.3$.  Thus the effect of the spatial
correlation is weakened because the correlation at distance $r$ is
described by the combination of $k_{c}r$, and the probability
$P(M_{1}\vert M_{2})$ is obtained by integrating $P(r,M_{1}\vert M_{2})$
from 0 to $R$ respect to $r$.  We checked this uncertainty, but the
effect does not change our results qualitatively, especially at larger
mass scale ($M\ga M_{*}$).

If we use a realistic power spectrum, for example CDM models, the
normalization of the mass spectrum $\sigma_{0}(M)$ becomes not
arbitrary.  In observation the normalization of the power spectrum is
given by the amplitude at $8h^{-1}$Mpc with the top-hat filter with the
bias parameter.  This leads that the amplitude at that scale does not
become unity in cases of other filters.  In Fig.\ref{fig:mf5}, we adopt
the LC method as a usual manner.  We have found that when we use the
simple relation the differences between the PS and the mass functions
with the Gaussian and the sharp $k$-space filters becomes large compared
with those in the case of the LC method.  Part of the remaining
differences is owing to the difference of the amplitude of the power
spectrum.

\subsection{Effect of non-spherical collapse}
Throughout this paper, we analyzed the mass function under the
simplification of the spherical collapse, in which the critical density
contrast for collapse is $\delta_{c}=1.69$.  However, in more realistic
condition, objects collapse nonspherically.  Monaco (1995; 1997a, b;
1998) showed that the {\it first} collapse, which means one-dimensional
or pancake collapse, occurs during more short time-scale.  Fitting the
PS mass function to the resultant mass function, he obtained an
effective density contrast $\delta_{c}\simeq 1.5$.  This predicts larger
number density of halos at larger scale compared to the PS, such as the
effect of the spatial correlation.  Apparently this seems to enhance the
difference of the mass function including the effects of the filtering,
peak, and spatial correlation.  It will be correct when we consider the
subsequent evolution of baryonic component included in halos because
shock heating of gas occurs at the first collapse.  Even so, in analyses
of the mass function by using $N$-body simulations which is in good
agreement with the theoretical PS mass function, it is not trivial that
the condition of the identification of isolated halos is corresponding
to the first collapse.  Actually such identified halos are rather
spherical.  This means that the effective critical density is larger
than 1.5 and that the second or third collapse criterion will be better
than the first collapse one.  If so, the non-spherical effect will
hardly affect our results, unless the mixture of the effects emerges as
shown in Fig.\ref{fig:mf4}.

\section{Summary and Conclusions}\label{sec:conc}
In this paper, we analyzed the following three effects on the mass
function: the filtering effect of the density fluctuation field, the
peak ansatz in which objects collapse around density maxima, and the
spatial correlation of density contrasts within a region of a collapsing
halo because objects have non-zero finite volume.  It has been shown by
many authors that the PS mass function agrees with $N$-body results
well.  However it had been also unclear why the PS mass function works
well.  Taking into account the above three effects and using an integral
formulation proposed by Jedamzik (1995), we tried to solve the missing
link in the PS formalism.  While YNG showed that the effect of the
spatial correlation alters the original PS mass function, in this paper,
we showed that the filtering effect almost cancels out the effect of the
spatial correlation and that the original PS is almost recovered by
combining these two effects, particularly in the case of the top-hat
filter.  As we shown in Fig.\ref{fig:mf5}, the mass function with this
filter is in good agreement with that given by a $N$-body simulation
(Jenkins et al. 2001).  On the other hand, as for the peak effect, the
resultant mass functions are changed dramatically and not in agreement
especially at low-mass tails in the cases of the Gaussian and the sharp
$k$-space filters.

In the sharp $k$-space filters, as shown by PH90 and BCEK, a density
contrast at a point changes as the smoothing scale decreases because a
new Fourier mode with an uncorrelated phase with other modes gives the
density contrast an additional value such as Markovian random walk,
which originates the so-called {\it fudge-factor-of-two}.  On the other
hand, when using general filters, we cannot expect that the trajectory
of density contrast against the smoothing scale behaves like Markovian
random walk.  Then the factor of two is lost.  We confirmed the result
by PH90 and BCEK by using the other formalism proposed by Jedamzik
(1995), against the filtering effect which we referred in this paper.
The number density of halos decreases at larger mass scale compared to a
typical mass scale which gives unity variance, and increase at smaller
mass scale.  Besides we found that the slope at smaller scale does not
change.  Thus apparently the shape of the multiplicity function $F(M)$
scales horizontally against the mass scale.  This effect can be
interpreted as the correlation of the Fourier modes, $\delta_{\bf k}$,
in a statistical sense.

We also investigated the effect of the peak ansatz in which objects
collapse around density maxima and which is often considered in
structure formation theories.  In the case of the Gaussian filter, this
makes the low-mass tail of the multiplicity function shallower than the
PS dramatically.  We found that this difference can be interpreted by
the behavior of the probability $P(M_{1}\vert M_{2})$ near $M_{1}\sim
M_{2}$.  $P(M_{1}\vert M_{2})$ increases above a half at that scale
while it remains a half in the sharp $k$-space filter.  On the other
hand, in the case of the top-hat filter, we cannot deal with this effect
correctly because this filter does not smooth out the density field
adequately so as to be able to differentiate the density contrasts
respect to the spatial coordinate due to the sharpness on the edge of
the filter.  While this trouble may be removed to adopt another filter
with a rounded edge, it will originate complicated mathematical
difficulty.

The effect of the spatial correlation only in the case of the sharp
$k$-space filter was already considered in YNG.  They showed that, in
contrast to the filtering effect shown here, the mass function predicts
more number of halos at larger mass scale, $M\ga M_{*}$, and less number
at smaller scale, $M\la M_{*}$.  This is apparently corresponding to
scaling the mass scale to larger one.  This property is caused by
decreasing correlation between density contrasts with different
smoothing scales.  By considering the other filters, we showed that such
properties are weakened and the resultant mass functions become close to
the original PS because the filtering effect strengths the mode
correlation as mentioned above.  We also showed, however, that the
spatial correlation cannot correct the change of the slope of the
low-mass tail in the case of the Gaussian filter which is caused by the
peak effect and besides steepens the tail in the case of the sharp
$k$-space filter.  This is because the probability $P(M_{1}\vert M_{2})$
is mainly affected at $M_{1}\sim M_{2}$ almost independent of the
absolute values of $M_{1}$ and $M_{2}$ by the peak effect, while mainly
affected at $M_{1}\ga M_{*}$ by the filtering effect and the spatial
correlation.  Probably we can confirm which filter is better by using a
high resolution $N$-body simulation, while it may be difficult to obtain
an adequate dynamic mass range to much smaller scale than $M_{*}$.

As a conclusion, the discrepancy with the PS shown by YNG, in which only
the sharp $k$-space filter was considered, is almost canceled out by the
filtering effect.  The peak ansatz, which does not affect the mass
function in the case of the top-hat filter due to the mathematical
reason, causes a new discrepancy at low mass scale in the cases of the
Gaussian and the sharp $k$-space filters.  A new filter similar to the
top-hat filter but differentiable will be required in order to consider
more realistic situation of collapsing halos.

It should be noted that the discrepancy is not perfectly canceled by the
filtering effect and the spatial correlation, especially at larger mass
scale, $M\ga M_{*}$.  This scale is corresponding to the galaxy
clusters.  Several authors have tried to determine the cosmological
parameters and the normalization factor of the power spectrum by
comparing the PS mass function to the observational cluster abundance
(e.g., Bahcall 2000).  We should pay attention to the fact that the
results by such approach may depend on the model mass function.

We discussed an important effect of the non-spherical collapse, which
has been investigated by Monaco (1995; 1997a, b; 1998).  He found that
if we adopt the first shell crossing as the collapse criterion, the
resultant mass function predicts larger number density of halos at
larger scale $M\ga M_{*}$ and smaller number density at smaller scale
$M\la M_{*}$ compared to the PS because the time-scale of collapse
decreases.  This effect will enhance the difference between the PS and
the mass function given by a {\it full} statistics with the effects
taken into account in this paper and with the non-spherical collapse.
Thus when we consider the evolution of the baryonic component, we should
pay attention this effect.  However, the agreement of the PS with the
mass function given by many $N$-body simulations will be obtained by
adopting the second or the third shell crossing as the collapse because
identified objects as halos are rather spherical, not sheet-like or
filamentary.

We found that the effects we considered becomes more prominent in larger
spectral index $n$.  In CDM model, the effective spectral index $n$ is
less than -2 at galactic scale.  Therefore in a realistic power spectrum
the difference of the low-mass tail of the mass function from the PS may
be negligible.  However, at larger scale $M\ga M_{*}$, the difference
remains.  This may leads a systematic error when the cosmological
parameters are estimated by comparing observations such as cluster
abundance with the PS.  Thus we need more detailed study on the mass
function.

Finally we should note that the approximation method used here is
different from the excursion set approach.  In the calculation of
$P(M_{1}\vert M_{2})$, we considered only two different mass scales,
$M_{1}$ and $M_{2}$.  However, more strictly speaking, the probability
must be subject to an additional condition, that is, the object with
$M_{2}$ is an {\it isolated} object.  In other words, the collapsing
condition of objects $M_{2}$ must describe the first upcrossing above
$\delta_{c}$.  This is essentially the same as the cloud-in-cloud
problem and the overlapping problem which was pointed out by YNG.  Even
so, we believe that the approximation used here is adequate to resolve
the mystery in the PS formalism because the fact that the filtering
effect correlates different Fourier modes leads to enhance the
probability that the objects $M_{2}$ in our formulation are isolated.
This conjecture can be stressed by the similarity of the mass functions
shown here and the Monte Carlo approach to solve the excursion set
formalism with the general filters, as mentioned above.  In future, this
will be proved by using a high resolution $N$-body simulation numerical
experimentally.

\acknowledgments 
I wish to thank N. Gouda, T. Yano, and F. Takahara for useful
suggestions.  Numerical computation in this work was partly carried out
at the Yukawa Institute Computer Facility.

\appendix 
\newenvironment{newenumerate}{
        \begin{enumerate}
        \renewcommand{\labelenumi}{\Roman{enumi}.}
        }{\end{enumerate}}
\newcommand{\sign}{\mbox{sign}}

\newcommand{\V}{{\bf V}}
\newcommand{\C}{{\bf C}}

\section{Formulation of $P(M_{1}\vert M_{2})$ for general filters}
In this paper, we only considered Gaussian random fields.  It is
well-known that the Gaussian field is perfectly determined by the
two-point correlation function.  In the followings, we show the form of
the probability required to analyze the mass function firstly, and show
the correlation and the auto-correlation functions, which are introduced
as covariances in multi-variate Gaussian distribution functions, in the
case of the power-law power spectrum, $P(k)\equiv\langle\vert\delta_{\bf
k}\vert^{2}\rangle=Ak^{n}$, where $A$ is the amplitude, which is
normalized so as to be unity at $M=M_{*}$, and the angle brackets stand
for the ensemble average.

The kernel in eq.(\ref{eqn:jed0}), $P(M_{1}\vert M_{2})$, is generally
expressed as a conditional probability.  Since we need at least two
Gaussian fields with smoothing scale $M_{1}$ and $M_{2}$ $(M_{1}\leq
M_{2})$, or $R_{1}$ and $R_{2}$ $(R_{1}\leq R_{2})$ in terms of length
as mentioned in \S 2, multi-variate Gaussian distribution functions are
required.  Here we define ${\bf V}^{N}\equiv (x_{1},\cdots, x_{N})$ as a
vector of $N$ Gaussian distributed variables and ${\bf C}^{N}$ as a
condition for $N$ variables.  The distribution function for a variable
$x_{N}$ under the condition ${\bf C}^{N-1}$ is written by Bayes'
theorem,
\begin{equation}
 p(x_{N}\vert \C^{N-1})d x_{N}=
\frac{p(x_{N}, \C^{N-1})d x_{N}}{p(\C^{N-1})},
\end{equation}
where $p(\V)$ is a multi-variate Gaussian distribution function.  In
this paper, the condition $\C^{N}$ means a criterion of collapse for
larger objects, $M_{2}$.  When we do not consider the peak condition,
the vector $\V$ is two-variable one, $\V^{2}=(\delta_{2}, \delta_{1})$
and the condition $\C$ constrains one variable $\delta_{2}$,
$\C^{1}=(\delta_{2}=\delta_{2c})$.  Note that the points for
$\delta_{1}$ and $\delta_{2}$ are generally not the same if the spatial
correlation is taken into account.  When we consider the peak condition,
eleven-variate Gaussian distribution is required as
$\V^{11}=(\delta_{2}, \eta_{i}, \zeta_{ij}, \delta_{1})$, where
$i=1,2,3$ and $i\leq j$.  Note that $\eta_{i}$ and $\zeta_{ij}$ are
derivatives of $\delta_{2}$ respect to the coordinate $x_{i}$.
Conditions for $\eta_{i}$ and $\zeta_{ij}$ are determined such that the
density has a local maximum at the point $M_{2}$.

By using a covariance matrix ${\bf M}$, the $N$-variate Guassian
distribution function is generally written as
\begin{equation}
 p(\V^{N})d\V^{N}=\frac{\exp[-Q/2]}{\sqrt{(2\pi)^{N}\det({\bf M})}}d\V^{N},
\end{equation}
where
\begin{eqnarray}
Q&=&\V{\bf M}^{-1}\V^{\rm T}\\
M_{ij}&=&\langle(x_{i}-\langle x_{i}\rangle)
 (x_{j}-\langle x_{j}\rangle)\rangle.
\end{eqnarray}
This shows that the multi-variate Gaussian distribution is perfectly
determined by correlations for any two variables.  Thus what we do is to
calculate the covariance matrix ${\bf M}$ for given variables.

As shown in YNG, the following covariances have non-zero values:
\begin{eqnarray}
\langle\delta_{1}^{2}\rangle&=&\sigma_{r}^{2},\qquad
\langle\delta_{2}^{2}\rangle=\sigma_{0}^{2},\qquad
\langle\delta_{1}\delta_{2}\rangle=\xi_{0},\qquad
\langle\eta_{i}\eta_{j}\rangle=\frac{\sigma_{1}^{2}}{3}\delta_{ij},\nonumber\\
\langle\zeta_{ij}\zeta_{kl}\rangle&=&\frac{\sigma_{2}^{2}}{15}
 (\delta_{ij}\delta_{kl}+\delta_{ik}\delta_{jl}+\delta_{il}\delta_{jk}),\qquad
\langle\delta_{2}\zeta_{ij}\rangle=-\frac{\sigma_{1}^{2}}{3}\delta_{ij},\qquad
\langle\delta_{1}\zeta_{ij}\rangle=-\frac{\xi_{1}}{3}\delta_{ij},
\end{eqnarray}
where $\delta_{ij}$ is the Kronecker's delta.  Variances and
correlations $\sigma_{r}, \sigma_{i}$, and $\xi_{i}$ are defined as
\begin{eqnarray}
\xi_{l}(r)&=&\frac{4\pi}{(2\pi)^{3}}\int_{0}^{\infty}
        \tW(kR_{1})\tW(kR_{2})P(k)\frac{\sin kr}{kr}k^{2l+2}dk
=\frac{4\pi A}{(2\pi)^{3}}\frac{1}{R_{2}^{n+2l+3}}I_{n+2l+2}(r),\\
\sigma_{l}^{2}&=&\frac{4\pi}{(2\pi)^{3}}\int_{0}^{\infty}
        \tW^{2}(kR_{2})P(k)k^{2l+2}dk
=\frac{4\pi A}{(2\pi)^{3}}\frac{1}{R_{2}^{n+2l+3}}I_{n+2l+2},\\
\sigma_{r}^{2}&=&\frac{4\pi}{(2\pi)^{3}}\int_{0}^{\infty}
        \tW^{2}(kR_{1})P(k)k^{2}dk
=\frac{4\pi A}{(2\pi)^{3}}\frac{1}{R_{1}^{n+3}}I_{n+2}.
\end{eqnarray}
Clearly $\xi_{l}(r)$ means the $2l$-th moment of the two-point
correlation function, $\sigma_{l}^{2}$ is the $2l$-th moment of the
variance with the smoothing scale $R_{2}$, and $\sigma_{r}^{2}$ is the
variance with the scale $R_{1}$.  The subscript $r$ stands for the value
with the scale $R_{1}$ because generally it is separated by $r$ from the
center of the object with the scale $R_{2}$ in this paper.  It should be
noted that $\xi_{l}(r)\to\sigma_{l}^{2}$ for $r\to 0$ and $R_{1}\to
R_{2}$.  The integrals $I_{m}(r)$ and $I_{m}$ are calculated for
specific window function $\tW(kR)$ in the next section.

In the followings, the peak condition is taken into account.  In the
case without the condition, the probabilities are obtained by neglecting
quantities concerning $\eta_{i}$ and $\zeta_{ij}$.  Here we diagonalize
the Hesse matrix $\zeta_{ij}$ by introducing the eigenvalues
$(-\lambda_{1}, -\lambda_{2}, -\lambda_{3})$ with the order
$\lambda_{1}\geq\lambda_{2}\geq\lambda_{3}$, and the Euler angles
$\alpha', \beta'$ and $\gamma'$.  The volume element concerning the
Hesse matrix becomes
\begin{equation}
 \prod_{\stackrel{\scriptstyle i,j=1,2,3}{i\leq j}}d\zeta_{ij}=
\vert(\lambda_{1}-\lambda_{2})(\lambda_{2}-\lambda_{3})
(\lambda_{3}-\lambda_{1})\vert d\lambda_{1}d\lambda_{2}d\lambda_{3}
\sin\beta' d\beta' d\alpha' d\gamma'.
\end{equation}
Here we normalize the variables,
\begin{equation}
 \nu_{1}\equiv\frac{\delta_{1}}{\sigma_{r}},
 \nu_{2}\equiv\frac{\delta_{2}}{\sigma_{0}},
 x'\equiv\frac{\lambda_{1}+\lambda_{2}+\lambda_{3}}{\sigma_{2}},
 y'\equiv\frac{\lambda_{1}-\lambda_{3}}{2\sigma_{2}},
 z'\equiv\frac{\lambda_{1}-2\lambda_{2}+\lambda_{3}}{2\sigma_{2}}.
\end{equation}
The peak condition constrains the derivatives of $\delta$ to be
$\bfeta=0$ and each diagonal component $-\lambda_{i}$ is less than 0,
which means $-y'\leq z'\leq y'$ and $y'\leq 0$.  The covariances are
transformed as
\begin{equation}
 \langle\nu_{1}^{2}\rangle=1,
 \langle\nu_{2}^{2}\rangle=1,
 \langle x'\nu_{2}\rangle=\gamma,
 \langle \nu_{1}\nu_{2}\rangle=\epsilon(r),
 \langle x'\nu_{1}\rangle=\mu(r),
 \langle x'^{2}\rangle=1,
 \langle y'^{2}\rangle=\frac{1}{15},
 \langle z'^{2}\rangle=\frac{1}{5},
\end{equation}
where $\epsilon(r)$, $\mu(r)$ and $\gamma$ are written as
\begin{eqnarray}
\epsilon(r)&=&\langle\nu_{1}\nu_{2}\rangle=\frac{\xi_{0}(r)}
        {\sigma_{r}\sigma_{0}},\\
\mu(r)&=&\langle x'\nu_{1}\rangle=\frac{\xi_{1}(r)}
        {\sigma_{r}\sigma_{2}},\\
\gamma&=&\langle x'\nu_{2}\rangle=\frac{\sigma_{1}^{2}}
        {\sigma_{0}\sigma_{2}},
\end{eqnarray}
and $\mu(r)\to\gamma$ for $r\to 0$ and $R_{1}\to R_{2}$.
In order to obtain the exponents of probabilities, we define the
following variables:
\begin{eqnarray}
Q_{a}&=&\frac{1-\gamma^{2}}{1-\epsilon^{2}-\mu^{2}-\gamma^{2}+
 2\epsilon\mu\gamma}
 \left(\nu_{1}+\frac{\mu\gamma-\epsilon}{1-\gamma^{2}}\nu_{2}
  +\frac{\epsilon\gamma-\mu}{1-\gamma^{2}}x'\right)^{2},\\
Q_{b}&=&\frac{(x'-\gamma\nu_{2})^{2}}{1-\gamma^{2}},\\
Q_{c}&=&\nu_{2}^{2}+15y'^{2}+5z'^{2}+
 3\frac{\vert\bfeta\vert^{2}}{\sigma_{1}^{2}},
\end{eqnarray}
where we omit the argument $r$ of $\epsilon(r)$ and $\mu(r)$ for
simplicity.  After integrating the volume element over the Euler angles,
we obtain
\begin{eqnarray} 
\lefteqn{p(\nu_{1},\nu_{2},\bfeta,x',y',z')d\nu_{1}d\nu_{2}d^{3}\bfeta dx'dy'dz'}\nonumber\\
&=&\frac{3(15)^{5/2}\vert y'(y'^{2}-z'^{2})\vert}
{8\sqrt{2}\pi^{7/2}\sigma_{1}^{3}
\sqrt{1-\epsilon^{2}-\mu^{2}-\gamma^{2}+2\epsilon\mu\gamma}}
\exp\left[-\frac{Q_{a}+Q_{b}+Q_{c}}{2}\right]
d\nu_{1}d\nu_{2}d^{3}\bfeta dx'dy'dz',\\
\lefteqn{p(\nu_{2},\bfeta,x',y',z')d\nu_{2}d^{3}\bfeta dx'dy'dz'}\nonumber\\
&=&\frac{3(15)^{5/2}\vert y'(y'^{2}-z'^{2})\vert}{8\pi^{3}\sigma_{1}^{3}
\sqrt{1-\gamma^{2}}}\exp\left[-\frac{Q_{b}+Q_{c}}{2}\right]
d\nu_{2}d^{3}\bfeta dx'dy'dz'.
\end{eqnarray}
The latter equation is the same as eq.(A7) in BBKS.  From the Bayes'
theorem, the required probability is written under the above condition
against $\nu_{2}, \bfeta, x', y'$ and $z'$ as
\begin{equation}
 P(r, M_{1}\vert M_{2})=\sqrt{\frac{1-\gamma^{2}}
  {2\pi (1-\epsilon^{2}-\mu^{2}-\gamma^{2}+2\epsilon\mu\gamma)}}
  \frac{\int_{0}^{\infty}dx' f(x')\int_{\nu_{1c}}^{\infty}d\nu_{1}
  e^{-(Q_{a}+Q_{b})/2}}
  {\int_{0}^{\infty}dx' f(x') e^{-Q_{b}/2}},\label{eqn:kernel}
\end{equation}
where
\begin{eqnarray}
f(x')&=&\left(\int_{0}^{x'/4}dy'\int_{-y'}^{y'}dz'+
	\int_{x'/4}^{x'/2}dy'\int_{3y'-x'}^{y'}dz'\right)
\vert y'(y'^{2}-z'^{2})\vert \exp\left(-\frac{15}{2}y'^{2}-
			       \frac{5}{2}z'^{2}\right)\nonumber\\
&=&\left(\int_{-x'/4}^{0}dz'\int_{-z'}^{(z'+x')/3}dy'+
	\int_{0}^{x'/2}dz'\int_{z'}^{(z'+x')/3}dy'\right)
[y'(y'^{2}-z'^{2})] \exp\left(-\frac{15}{2}y'^{2}-
			       \frac{5}{2}z'^{2}\right)\nonumber\\
&=&\frac{1}{2250}\left[-15x'\exp\left(-\frac{5}{8}x'^{2}\right)
	+\sqrt{10\pi}\left\{
	{\rm erf}\left(\frac{x'}{2}\sqrt{\frac{5}{2}}\right)+
	{\rm erf}\left(x'\sqrt{\frac{5}{2}}\right)
		\right\}\right],
\end{eqnarray}
with the definition of the error function,
\begin{equation}
 {\rm erf}(x)=\frac{2}{\sqrt{\pi}}\int_{0}^{x}e^{-t^{2}}dt.
\end{equation}
Note that the form of $f(x)$ is different from eq.(A15) in BBKS because
they considered not the probability but the number density of peaks.
Finally, if we take into account the spatial correlation, that is, the
different points of $\delta_{1}$ and $\delta_{2}$, we need to spatially
average the above probability over the region $R_{2}$ in order to obtain
$P(M_{1}\vert M_{2})$,
\begin{equation}
 P(M_{1}\vert M_{2})=\frac{\int_{0}^{R}P(r, M_{1}\vert M_{2})4\pi r^{2}dr}
  {\int_{0}^{R}4\pi r^{2}dr}.
\end{equation}

\section{Correlation coefficients for specific filters}
In the followings, we give the specific forms of the correlation
coefficients $\epsilon(r), \mu(r)$ and $\gamma$ for the three filters,
the top-hat, the Gaussian, and the sharp $k$-space filters, and for the
spectral index $n=0$ and $-2$, which are important values when
considering formation process of galaxies and galaxy clusters in a CDM
universe.

Here, for convenience, we define the following dimensionless variables,
\begin{equation}
y=kR_{2},\quad z=R_{1}/R_{2}, \quad s=r/R_{2}.
\end{equation}
By using these variables, the correlation coefficients are written as
follows:
\begin{equation}
\gamma=\frac{I_{n+4}}{\sqrt{I_{n+2}I_{n+6}}},\qquad
\epsilon(r)=z^{\frac{n+3}{2}}\frac{I_{n+2}(r)}{I_{n+2}},\qquad
\mu(r)=z^{\frac{n+3}{2}}\frac{I_{n+4}(r)}{\sqrt{I_{n+2}I_{n+6}}}.
\end{equation}
Thus, the problem is reduced to calculating the integrals $I_{m}(r)$ and
$I_{m}$.

\subsection{Top-hat filter}
In the case of the top-hat filter, the behavior of the correlations
changes at the boundary of an object with the smoothing scale $R_{2}$.
So we define the regions I, II, and III as 
\begin{newenumerate}
\item Internal: $r\leq R_{2}-R_{1}$,
\item Crossing: $R_{2}-R_{1}\leq r\leq R_{2}+R_{1}$,
\item External: $R_{2}+R_{1}\leq r$,
\end{newenumerate}
and the following new variables as 
\begin{equation}
a=1-z-s, \quad b=1-z+s, \quad c=1+z-s, \quad d=1+z+s.
\end{equation}
Actually the region III is not considered in our analyses, but we
express the explicit forms of the correlations for the region III in the
following.

The generic forms of the integrals can be expressed as
\begin{eqnarray}
I_{m}(r)&=&\int_{0}^{\infty}\frac{9}{y^{6}z^{3}}(\sin yz-yz\cos yz)
(\sin y-y\cos y)y^{m}\frac{\sin ys}{ys}dy,\\
I_{m}&=&\int_{0}^{\infty}\frac{9}{y^{6}}(\sin y-y\cos y)^{2}y^{m}dy.
\end{eqnarray}
The solutions for $m=0, 2, 4$ and 6 depend on the region I, II and III
as follows:
\begin{eqnarray}
I_{0}(r)=\frac{\pi}{640sz^{3}}[
&-&6\vert a\vert (1-z)a^{4}+\sign(a)(a^{6}-30za^{4})\nonumber\\
&+&6\vert b\vert (1-z)b^{4}+\sign(b)(-b^{6}+30zb^{4})\nonumber\\
&+&6\vert c\vert (1+z)c^{4}+\sign(c)(-c^{6}-30zc^{4})\nonumber\\
&-&6\vert d\vert (1+z)d^{4}+\sign(d)(d^{6}+30zd^{4})],
\end{eqnarray}
\begin{eqnarray}
I_{2}(r)=\frac{9\pi}{192sz^{3}}[
&&4\vert a\vert (1-z)a^{2}+\sign(a)(-a^{4}+12za^{2})\nonumber\\
&-&4\vert b\vert (1-z)b^{2}+\sign(b)(b^{4}-12zb^{2})\nonumber\\
&-&4\vert c\vert (1+z)c^{2}+\sign(c)(c^{4}+12zc^{2})\nonumber\\
&+&4\vert d\vert (1+z)d^{2}+\sign(d)(-d^{4}-12zd^{2})],
\end{eqnarray}
\begin{eqnarray}
I_{4}(r)=\frac{9\pi}{16sz^{3}}[
&-&2\vert a\vert (1-z)+\sign(a)(a^{2}-2z)\nonumber\\
&+&2\vert b\vert (1-z)+\sign(b)(-b^{2}+2z)\nonumber\\
&+&2\vert c\vert (1+z)+\sign(c)(-c^{2}-2z)\nonumber\\
&-&2\vert d\vert (1+z)+\sign(d)(d^{2}+2z)],
\end{eqnarray}
\begin{equation}
I_{6}(r)=\frac{9\pi}{8sz^{3}}[-\sign(a)+\sign(b)+\sign(c)-\sign(d)],
\end{equation}
or more explicitly,
\begin{eqnarray}
I_{0}(r)&=&
\left\{
\begin{array}{ll}
\displaystyle{
\frac{\pi}{20}(-3z^{2}+15-5s^{2})}&\qquad\mbox{I}\\
\displaystyle{
\frac{\pi}{640sz^{3}}[10(z^{6}-9z^{4}+16z^{3}-9z^{2}+1)+
48s(-z^{5}+5z^{3}+5z^{2}-1)}&\\
\displaystyle{\quad
+90s^{2}(z^{4}-2z^{2}+1)+80s^{3}(-z^{3}-1)
+30s^{4}(z^{4}+1)-2s^{6}]}&\qquad\mbox{II}\\
\displaystyle{\frac{\pi}{2s}}&\qquad\mbox{III},
\end{array}\right.\\
I_{2}(r)&=&
\left\{
\begin{array}{ll}
\displaystyle{\frac{3\pi}{2}}&\qquad\mbox{I}\\
\displaystyle{
\frac{3\pi}{32sz^{3}}[-3(1-z^{2})^{2}+8s(1+z^{3})-6s^{2}(1+z^{2})+s^{4}]
}&\qquad\mbox{II}\\
\displaystyle{0}&\qquad\mbox{III},
\end{array}\right.\\
I_{4}(r)&=&
\left\{
\begin{array}{ll}
\displaystyle{0}&\qquad\mbox{I}\\
\displaystyle{\frac{9\pi}{8sz^{3}}(z^{2}-s^{2}+1)}&\qquad\mbox{II}\\
\displaystyle{0}&\qquad\mbox{III},
\end{array}\right.\\
I_{6}(r)&=&
\left\{
\begin{array}{ll}
\displaystyle{0}&\qquad\mbox{I}\\
\displaystyle{\frac{9\pi}{4sz^{3}}}&\qquad\mbox{II}\\
\displaystyle{0}&\qquad\mbox{III},
\end{array}\right.
\end{eqnarray}
and
\begin{eqnarray}
I_{0}&=&\frac{3\pi}{5},\\
I_{2}&=&\frac{3\pi}{2},\\
I_{4}&=&\lim_{y\to\infty}\frac{9}{4y}[-2+2y^{2}+2\cos{2y}+y\sin{2y}]\to\infty,\\
I_{6}&=&\lim_{y\to\infty}\frac{9}{24}[12y+4y^{3}+18y\cos{2y}-
15\sin{2y}+6y^{2}\sin{2y}]\to\infty.
\end{eqnarray}
The correlation coefficients are obtained by using the above integrals
for the spectral index $n=0$,
\begin{eqnarray}
\gamma&=&0,\\
\epsilon(r)&=&z^{3/2}\frac{I_{2}(r)}{3\pi/2}=
\left\{
\begin{array}{ll}
\displaystyle{z^{3/2}}&\qquad\mbox{I}\\
\displaystyle{\frac{z^{-3/2}}{16s}(-3(1-z^{2})^{2}+8s(1+z^{3})
-6s^{2}(1+z^{2})+s^{4})}&\qquad\mbox{II}\\
\displaystyle{0}&\qquad\mbox{III},
\end{array}\right.\\
\mu(r)&=&z^{3/2}\frac{I_{4}(r)}{\sqrt{I_{2}I_{6}}}\to0,
\end{eqnarray}
and for $n=-2$,
\begin{eqnarray}
\gamma&=&0,\\
\epsilon(r)&=&z^{1/2}\frac{
I_{0}(r)}{3\pi/5}
=\left\{
\begin{array}{ll}
\displaystyle{\frac{z^{1/2}}{12}(-3z^{2}+15-5s^{2})}&\qquad\mbox{I}\\
\displaystyle{\frac{z^{-5/2}}{384s}[10(z^{6}-9z^{4}+16z^{3}-9z^{2}+1)
+48s(-z^{5}+5z^{3}+5z^{2}-1)}\\
\qquad\displaystyle{+90s^{2}(z^{4}-2z^{2}+1)-80s^{3}(z^{3}+1)
+30s^{4}(z^{2}+1)-2s^{6}]}&\qquad\mbox{II}\\
\displaystyle{\frac{5z^{1/2}}{6s}}&\qquad\mbox{III},
\end{array}\right.\\
\mu(r)&=&z^{1/2}\frac{I_{2}(r)}{\sqrt{I_{0}I_{4}}}\to0.
\end{eqnarray}
These show that we cannot deal with the configuration of density peaks,
at least statistically, in the case of the top-hat filter because of the
divergence of $I_{4}$ and $I_{6}$ to infinity.  This may be originated
by the sharpness of the window function, which do not adequately smooth
out density fields so as to be differentiable.

Here we show that the peak condition is canceled out in the calculation
of the probability $P(r,M_{1}\vert M_{2})$.  Now $\mu(r)=\gamma=0$, then
the eq.(\ref{eqn:kernel}) becomes
\begin{equation}
P(r,M_{1}\vert M_{2})=\sqrt{\frac{1}{2\pi(1-\epsilon^{2})}}\frac{
\int_{0}^{\infty}dx'f(x')
\int_{\nu_{1c}}^{\infty}d\nu_{1}e^{-\frac{Q_{a}+Q_{b}}{2}}}
{\int_{0}^{\infty}dx'f(x')e^{-\frac{Q_{b}}{2}}},
\end{equation}
and the exponents become
\begin{equation}
Q_{a}=\frac{(\nu_{1}-\epsilon\nu_{2c})^{2}}{1-\epsilon^{2}},\qquad
Q_{b}=x'^{2}.
\end{equation}
Because the exponent $Q_{a}$ does not depend on $x'$, we can cancel the
integration respect to $x'$ as follows,
\begin{equation}
P(r,M_{1}\vert M_{2})=\frac{1}{\sqrt{2\pi(1-\epsilon^{2})}}\int_{\nu_{1c}}^{\infty}d\nu_{1}e^{-\frac{Q_{a}}{2}}.
\end{equation}
This is the same form as the probability not taking into account the
peak effect.

\subsection{Gaussian filter}
The generic forms of the integrals are expressed by using the confluent
hypergeometric function and the gamma function as follows:
\begin{eqnarray}
I_{m}(r)&=&\int_{0}^{\infty}e^{-\frac{y^{2}}{2}(1+z^{2})}y^{m}
\frac{\sin ys}{ys}dy
=\frac{\Gamma(\frac{1+m}{2})}{[2(1+z^{2})]^{\frac{1+m}{2}}}
{}_{1}F_{1}\left(\frac{1+m}{2};\frac{3}{2};-\frac{s^{2}}{2(1+z^{2})}\right),\\
I_{m}&=&\int_{0}^{\infty}e^{-y^{2}}y^{m}dy.
\end{eqnarray}
Specifically,
\begin{eqnarray}
I_{0}(r)&=&\frac{\pi}{2s}\mbox{erf}\left[\frac{s}{\sqrt{2(1+z^{2})}}\right],\\
I_{2}(r)&=&\sqrt{\frac{\pi}{2}}\frac{1}{(1+z^{2})^{3/2}}
e^{-\frac{s^{2}}{2(1+z^{2})}},\\
I_{4}(r)&=&\sqrt{\frac{\pi}{2}}\frac{3}{(1+z^{2})^{5/2}}
\left[1-\frac{s^{2}}{3(1+z^{2})}\right]e^{-\frac{s^{2}}{2(1+z^{2})}},\\
I_{6}(r)&=&\sqrt{\frac{\pi}{2}}\frac{15}{(1+z^{2})^{7/2}}
\left[1-\frac{2s^{2}}{3(1+z^{2})}+\frac{s^{4}}{15(1+z^{2})^{2}}\right]
e^{-\frac{s^{2}}{2(1+z^{2})}},
\end{eqnarray}
and
\begin{equation}
I_{m}=\frac{\Gamma(\frac{1+m}{2})}{2},\quad
I_{0}=\frac{\sqrt{\pi}}{2},\quad
I_{2}=\frac{\sqrt{\pi}}{4},\quad
I_{4}=\frac{3\sqrt{\pi}}{8},\quad
I_{6}=\frac{15\sqrt{\pi}}{16}.
\end{equation}
The correlation coefficients for $n=0$ are
\begin{eqnarray}
\gamma&=&\sqrt{\frac{3}{5}},\\
\epsilon(r)&=&\left(\frac{2z}{1+z^{2}}\right)^{3/2}e^{-\frac{s^{2}}{2(1+z^{2})}},\\
\mu(r)&=&4\sqrt{\frac{6}{5}}\frac{z^{3/2}}{(1+z^{2})^{5/2}}
\left[1-\frac{s^{2}}{3(1+z^{2})}\right]e^{-\frac{s^{2}}{2(1+z^{2})}},
\end{eqnarray}
and for $n=-2$,
\begin{eqnarray}
\gamma&=&\frac{1}{\sqrt{3}},\\
\epsilon(r)&=&\frac{\sqrt{\pi z}}{s}{\rm erf}
\left[\frac{s}{\sqrt{2(1+z^{2})}}\right],\\
\mu(r)&=&\sqrt{\frac{8}{3}}\frac{z^{1/2}}{(1+z^{2})^{3/2}}e^{-\frac{s^{2}}{2(1+z^{2})}}.
\end{eqnarray}

\subsection{Sharp $k$-space filter}
Here we show the integrals by using the generic cut-off wavenumber
$k_{c}$ which is introduced in the filter, $y_{c}=k_{c}R$.  In this
paper we showed the results for $y_{c}=\pi$.  If the definition by the
LC method is used, $y_{c}=(9\pi/2)^{1/3}$.  The generic forms of the
integrals are
\begin{eqnarray}
I_{m}(r)&=&\int_{0}^{y_{c}}y^{m}\frac{\sin{ys}}{ys}dy=
\frac{y_{c}^{1+m}}{1+m}{}_{1}F_{2}\left[\frac{1+m}{2};
\frac{3}{2},\frac{3+m}{2};-\frac{(sy_{c})^{2}}{4}\right],\\
I_{m}&=&y_{c}^{m+1},
\end{eqnarray}
where ${}_{1}F_{2}(a;b,c;z)$ is the generalized hypergeometric function.
The specific forms of $I_{m}(r)$ are
\begin{eqnarray}
I_{0}(r)&=&\frac{1}{s}\int_{0}^{sy_{c}}\frac{\sin{t}}{t}dt,\\
I_{2}(r)&=&\frac{1}{s^{3}}[\sin{sy_{c}}-sy_{c}\cos{sy_{c}}],\\
I_{4}(r)&=&\frac{1}{s^{5}}[sy_{c}(6-s^{2}y_{c}^{2})\cos{sy_{c}}
-3(2-s^{2}y_{c}^{2})\sin{sy_{c}}],\\
I_{6}(r)&=&\frac{1}{s^{7}}[-sy_{c}(120-20s^{2}y_{c}^{2}+s^{4}y_{c}^{4})
\cos{sy_{c}}+5(24-12s^{2}y_{c}^{2}+s^{4}y_{c}^{4})\sin{sy_{c}}].
\end{eqnarray}
The correlation coefficients for $n=0$ are
\begin{eqnarray}
\gamma&=&\frac{\sqrt{21}}{5},\\
\epsilon(r)&=&\frac{3z^{3/2}}{s^{3}y_{c}^{3}}[\sin{sy_{c}}-sy_{c}\cos{sy_{c}}],\\
\mu(r)&=&\frac{\sqrt{21}z^{3/2}}{s^{5}y_{c}^{5}}[sy_{c}(6-s^{2}y_{c}^{2})
\cos{sy_{c}}-3(2-s^{2}y_{c}^{2})\sin{sy_{c}}],
\end{eqnarray}
and for $n=-2$,
\begin{eqnarray}
\gamma&=&\frac{\sqrt{5}}{3},\\
\epsilon(r)&=&\frac{z^{1/2}}{sy_{c}}\int_{0}^{sy_{c}}\frac{\sin{t}}{t}dt,\\
\mu(r)&=&\frac{\sqrt{5}z^{1/2}}{s^{3}y_{c}^{3}}[\sin{sy_{c}}-sy_{c}\cos{sy_{c}}].
\end{eqnarray}


\begin{thebibliography}{}   
\bibitem{}Appel, A., \& Jones, B.T. 1990, MNRAS, 245, 522
\bibitem{}Bahcall, N.A. 2000, \physrep, 333, 233
\bibitem{}Bardeen, J.M., Bond, J.R., Kaiser, N., \& Szalay, A.S. 1986,
	ApJ, 304, 15 (BBKS)
\bibitem{}Bond, J.R., Cole, S., Efstathiou, G., \& Kaiser, N. 1991, ApJ,
	379, 440 (BCEK)
\bibitem{}Bower, R. 1991, MNRAS, 248, 332
\bibitem{}Doroshkevich, A.G. 1970, Astrofizika, 6, 581
	(trans. Astrophysics, 6, 320 [1973])
\bibitem{}Epstein, R.I. 1983, MNRAS, 205,207   
\bibitem{}Epstein, R.I. 1984, ApJ, 281, 545
\bibitem{}Jedamzik, K. 1995, ApJ, 448, 1   
\bibitem{}Jenkins, A. et al. 2001, MNRAS, 321, 372
\bibitem{}Gunn, J.E., \& Gott, J.R. 1972, ApJ, 176, 1
\bibitem{}Kauffmann, G., \& White, S.D.M. 1993, MNRAS, 261, 921
\bibitem{}Lacey, C.G., \& Cole, S. 1993, MNRAS, 262, 627 (LC)
\bibitem{}Manrique, A., \& Salvador-Sol{\'e}, E. 1995, ApJ, 453, 6
\bibitem{}Monaco, P. 1995, ApJ, 447, 23
\bibitem{}Monaco, P. 1997a, MNRAS, 287, 753
\bibitem{}Monaco, P. 1997b, MNRAS, 290, 439
\bibitem{}Monaco, P. 1998, Fundam. Cosmic Phys., 19, 157
\bibitem{}Nagashima, M., \& Gouda, N. 1997, MNRAS, 287, 515
\bibitem{}Peacock, J.A., \& Heavens, A.F. 1985, MNRAS, 217, 805   
\bibitem{}Peacock, J.A., \& Heavens, A.F. 1990, MNRAS, 243, 133 (PH90)
\bibitem{}Press, W.H., \& Schechter, P. 1974, ApJ, 187, 425 (PS)
\bibitem{}Rodrigues, D.D.C., \& Thomas, P.A. 1996, MNRAS, 282, 631
\bibitem{}Sheth, R.K., \& Tormen, G. 1999, MNRAS, 308, 119
\bibitem{}Somerville, R.S., \& Kolatt, T. 1998, MNRAS, 305, 1
\bibitem{}Tomita, K. 1969, PTP, 42, 9
\bibitem{}Yano, T., Nagashima, M., \& Gouda, N. 1996, ApJ, 466, 1 (YNG)
\bibitem{}Zel'dovich, Ya.B. 1970, Astrofizika, 6,319
	(trans. Astrophysics, 6, 164 [1973])
\end{thebibliography}
\end{document}